\documentstyle[psfig,macros]{mn}
\begin{document}
%

\title[Coming of Age in the Dark Sector]
      {Coming of Age in the Dark Sector: How Dark Matter Haloes grow their Gravitational Potential Wells}

\author[van den Bosch et al.]
       {\parbox[t]{\textwidth}{
        Frank C. van den Bosch$^{1}$\thanks{E-mail: frank.vandenbosch@yale.edu}, 
        Fangzhou Jiang$^{1}$, 
        Andrew Hearin$^{2}$,
        Duncan Campbell$^1$, \\
        Douglas Watson$^{3}$\thanks{NSF Astronomy \& Astrophysics Postdoctoral Fellow},
        Nikhil Padmanabhan$^2$} \\ 
           \vspace*{3pt} \\
        $^1$Department of Astronomy, Yale University, PO. Box 208101,
            New Haven, CT 06520-8101\\
        $^2$Department of Physics, Yale University, 260 Whitney Avenue, 
            New Haven, CT 06520\\
        $^3$Kavli Institute for Cosmological Physics, University of
            Chicago, 933 East 56$^{\rm th}$ Street, Chicago, IL 60637}


\date{}

\pagerange{\pageref{firstpage}--\pageref{lastpage}}
\pubyear{2013}

\maketitle

\label{firstpage}


\begin{abstract}
  We present a detailed study of how dark matter haloes assemble their
  mass and grow their (central) potential well. We characterize these
  via their mass accretion histories (MAHs) and potential well growth
  histories (PWGHs), which we extract from the Bolshoi simulation and
  from semi-analytical merger trees supplemented with a method to
  compute the maximum circular velocity, $\Vmax$, of progenitor
  haloes. The results of both methods are in excellent agreement, both
  in terms of the average and the scatter. We show that the MAH and
  PWGH are tightly correlated, and that growth of the central
  potential precedes the assembly of mass; the maximum circular
  velocity is already half the present day value by the time the halo
  has accreted only 2 percent of its final mass. Finally, we
  demonstrate that MAHs have a universal form, which we use to develop
  a new and improved universal model that can be used to compute the
  average or median MAH and PWGH for a halo of any mass in any
  $\Lambda$CDM cosmology, without having to run a numerical simulation
  or a set of halo merger trees.
\end{abstract} 




\section{Introduction} 
\label{sec:intro}

In the current Cold Dark Matter (CDM) paradigm, the hierarchical
growth and buildup of dark matter haloes is the foundation on which
galaxy formation and evolution unfolds itself. Hence, the
understanding and characterization of the formation and properties of
CDM haloes is a fundamental component of any theory of galaxy
formation (see Mo \etal 2010 for a detailed overview).

The formation history of a dark matter halo is conveniently
characterized by its merger tree, which describes how its progenitors
merge and accrete over cosmic time. For a given cosmological model,
such merger trees can be constructed either from $N$-body simulations
or from Monte-Carlo realizations based on the extended Press-Schechter
(EPS) formalism (Bond \etal 1991; Bower 1991; Lacey \& Cole
1993). When tracing a merger tree back in time, each halo breaks up
into progenitor haloes, which themselves break up into progenitors,
etc.  The most massive progenitor of a given descendant halo (or the
progenitor that contributes most mass to its descendant, see
\S\ref{sec:SIM}) is called its {\it main} progenitor, and the {\it
  main branch} of the merger tree is defined as the branch tracing the
main progenitor of the main progenitor of the main progenitor,
etc. The mass history, $M(z)$, along this main branch is called the
{\it mass accretion history} (hereafter MAH)\footnote{Also sometimes
  called the `mass assembly history' or `mass aggregation history'.},
and plays a crucial role; it is the property that is most often used
to define halo formation (or assembly) times (e.g., Lacey \& Cole
1993; Nusser \& Sheth 1999; van den Bosch 2002a; Neistein, van den
Bosch \& Dekel 2006; Li \etal 2007, 2008; Giocoli et al. 2007;
Giocoli, Tormen \& Sheth 2012), and it is strongly correlated with
various structural properties of dark matter haloes, such as halo
concentration (e.g., Wechsler \etal 2002; Zhao \etal 2003b, 2009;
Tasitsiomi \etal 2004; Giocoli \etal 2012; Ludlow \etal 2013), and
substructure abundance (e.g., Gao \etal 2004; van den Bosch, Tormen \&
Giocoli 2005; Giocoli \etal 2010; Yang \etal 2011; Jiang \& van den
Bosch 2014b). In addition, the time derivative of the MAH gives the
mass growth rate, which is directly related to the rate at which
haloes accrete baryons from the cosmic web (e.g., van den Bosch 2002a;
Neistein \& Dekel 2008a,b; Genel \etal 2009; McBride, Fakhouri \& Ma
2009; Fakhouri, Ma \& Boylan-Kolchin 2010), which is why MAHs are
often used as the backbone for modeling the formation of (disk)
galaxies (e.g., Eisenstein \& Loeb 1996; Avila-Reese, Firmani \&
Hern\'andez 1998; Firmani \& Avila-Reese 2000; van den Bosch 2001,
2002b; Dutton \etal 2007). In fact, many properties of present-day
galaxies are expected to be tightly correlated with the MAH of their
host halo, including their morphology (e.g., Kauffmann, White \&
Guiderdoni 1993; Baugh, Cole \& Frenk 1996) and their star formation
rate (e.g., Hearin \& Watson 2013; Watson \etal 2014; Hearin, Watson
\& van den Bosch 2014).

Numerous studies have investigated the properties of MAHs in
($\Lambda$)CDM cosmologies, which have revealed a number of
trends. Using high resolution $N$-body simulations, Zhao \etal
(2003a,b) found that the MAHs typically consist of two distinct
phases: an early phase of rapid mass growth, and a later phase of slow
growth (see also Tasitsiomi \etal 2004; Li \etal 2007). The early,
rapid growth phase is characterized by major mergers which keep the
system from settling in virial equilibrium (e.g., Cohn \& White
2005). During this period the halo continuously reconfigures its
structure through violent relaxation and phase mixing, while deepening
its potential well. For reasons that are not yet understood, the
end-point of this process is a (now virialized) dark matter halo with
an NFW density profile (Navarro, Frenk \& White 1997) with a
concentration parameter, $c \sim 4$ (see \S\ref{sec:def} below for a
definition of $c$). During the subsequent slow growth phase the halo
predominantly grows in mass due to minor mergers and the accretion of
diffuse matter. This mainly adds matter to the outskirts, and
therefore causes the halo concentration parameter to increase (e.g.,
Zhao \etal 2009).

van den Bosch (2002a; hereafter vdB02) used EPS merger trees and
numerical simulations to show that the {\it average} MAHs (averaged
over many haloes of given mass), follow a simple, universal profile
that is characterized by two parameters that scale with halo mass and
cosmology. In particular, more massive haloes were shown to assemble
later, which gives rise to a decreasing concentration-mass relation,
and which is a simple consequence of the fact that the mass variance
of the CDM power spectrum decreases with increasing mass. The
halo-to-halo variance of MAHs, even for haloes of fixed present-day
mass, is large. It is responsible for the scatter in the
concentration-mass relation (e.g., Wechsler \etal 2002), is one of the
dominant sources of scatter in the Tully-Fisher relation (e.g.,
Eisenstein \& Loeb 1996; van den Bosch 2000; Dutton \etal 2007), and
is correlated with the halo's large scale environment (e.g., Sheth \&
Tormen 2004; Gao, Springel \& White 2005; Harker \etal 2006;
Maulbetsch \etal 2007).

Whereas the MAHs of dark matter haloes have received abundant
attention in the literature, there has been surprisingly little focus
on the buildup of dark matter potential wells. In particular, the
central potential depth is an important property. As shown in
\S\ref{sec:def}, it is directly proportional to the halo's maximum
circular velocity, $\Vmax$, which is often used as the halo parameter
of choice in abundance matching (e.g., Trujillo-Gomez \etal 2011;
Reddick \etal 2012; Hearin \etal 2013; Zentner, Hearin \& van den
Bosch 2014). This suggests that $\Vmax$ may be a better `regulator' of
galaxy formation than halo mass. This should not come as a surprise,
given that feedback is a crucial ingredient of galaxy formation, and
the efficiency of feedback processes to expel matter and metals
depends on the escape speed, and hence the depth of the central
potential well. $\Vmax$ also has the advantage that it can be much
more reliably and robustly measured than halo mass, both from
simulations and in observational data. In addition, it is free from
issues related to `pseudo-evolution' (Diemand, Kuhlen \& Madau 2007;
Cuesta \etal 2008; Zemp 2013; Diemer, More \& Kravtsov 2013; Diemer \&
Kravtsov 2014), and is defined without any ambiguity, unlike halo
mass, for which multiple definitions are in use\footnote{to great
  frustration of most practitioners, who constantly have to convert
  one definition to another.}.

The main goal of this paper is to study the potential well growth
histories (hereafter PWGHs) of CDM haloes, which we characterize as
the temporal evolution of $\Vmax$ of a halo's main progenitor. In
particular, we aim to provide a `recipe' to compute the average PWGH
for a halo of any given mass in any (reasonable) cosmology. To do so,
we use EPS-based merger trees combined with a model, introduced in
Jiang \& van den Bosch (2014b), that allows us to compute $\Vmax$ for
each progenitor halo along the tree. We test and calibrate our model
using merger trees and PWGHs extracted from the Bolshoi simulation
(Klypin, Trujillo-Gomez \& Primack 2011), and show that the PWGH can
be inferred from the MAH combined with a model for the
concentration-mass-redshift relation of dark matter
haloes. Unfortunately, previous fitting functions used to describe
average or median halo MAHs are inadequate, in that they either use a
different definition for the main progenitor, or they are only valid
for a single cosmology. We instead provide a new, universal model, and
demonstrate that the average (and median) MAHs of haloes of different
mass and in different cosmologies are related via a simple time
transformation that is motivated by insights gained from the EPS
formalism (see also Neistein, Macc\'io \& Dekel 2010).

This paper is organized as follows. In \S\ref{sec:Method} we give a
brief overview of the basics of dark matter haloes, followed by
descriptions of our semi-analytical model and of the numerical
simulation used to calibrate and test the model. In \S\ref{sec:res} we
present the MAHs and PWGHs of dark matter haloes obtained using our
semi-analytical model, and compare them to results from the Bolshoi
simulation and to predictions from the models by Zhao \etal (2009) and
Giocoli \etal (2012). In \S\ref{sec:uni} we present our new and
improved universal model for the average and median MAHs of dark
matter haloes. We show how it can be used to compute the corresponding
PWGHs, and use it to derive a fully analytical model for the average
mass accretion rates of dark matter haloes. We summarize our findings
in \S\ref{sec:conclusion}.


\section{Methodology}
\label{sec:Method}

This section describes the numerical simulations and semi-analytical
models used to study the Mass Accretion Histories (MAHs) and Potential
Well Growth Histories (PWGHs) of dark matter haloes. However, we start
with a brief introduction of halo basics, outlining a number of
definitions and notations.

\subsection{Halo Basics and Notation}
\label{sec:def}

Throughout this paper dark matter haloes at redshift $z$ are defined
as spherical systems with a virial radius $r_{\rm vir}$ inside of
which the average density is equal to $\Delta_{\rm vir}(z) \,
\rho_{\rm crit}(z)$. Here $\rho_{\rm crit}(z) = 3 H^2(z)/8 \pi G$ is
the critical density for closure, and is given by
\begin{equation}\label{deltavir}
\Delta_{\rm vir}(z) = 18\pi^2 + 82 x - 39 x^2\,
\end{equation}
with $x =\Omega_\rmm(z) - 1$ (Bryan \& Norman 1998).  The (virial)
mass of a dark matter halo is defined as the mass within the virial
radius $r_{\rm vir}$ and indicated by $M$.

We also assume throughout that dark matter haloes follow an NFW
density profile
\begin{equation}\label{rhoNFW}
\rho(r) = \rho_{\rm crit} {\delta_{\rm char} \over (r/r_\rms)\, 
(1 + r/r_\rms)^2} \,.
\end{equation}
Here $r_\rms$ is the scale radius, and 
\begin{equation}\label{deltachar}
\delta_{\rm char} = {\Delta_{\rm vir} \over 3} \, {c^3 \over f(c)} \,,
\end{equation}
with $c = r_{\rm vir}/r_\rms$ the halo concentration parameter, and
\begin{equation}\label{fx}
f(x) = \ln(1+x) - {x \over 1+x}\,.
\end{equation}
The maximum circular velocity of a NFW halo occurs at a radius 
$r_{\rm max} \simeq 2.16\,r_\rms$, and is given by
\begin{equation} \label{VmaxHost}
V_{\rm max} = 0.465 \, V_{\rm vir} \, \sqrt{c \over f(c)}\,,
\end{equation}
where 
\begin{eqnarray}\label{Vvir}
\lefteqn{V_{\rm vir} = 159.43 \kms \, \left({M \over 10^{12}\msunh}\right)^{1/3}
\, \left[{H(z) \over H_0}\right]^{1/3}} \nonumber \\
& & \, \left[{\Delta_{\rm vir}(z) \over 178}\right]^{1/6}\,,
\end{eqnarray}
is the virial velocity, defined as the circular velocity at the virial
radius. The gravitational potential of a spherical NFW density
distribution is given by
\begin{equation}
\Phi(r) = -V^2_{\rm vir} \, {\ln(1+cx) \over f(c) \, x} =
- \left({V_{\rm max} \over 0.465}\right)^2 \, {\ln(1+cx) \over cx}\,,
\end{equation}
where $x=r/r_{\rm vir}$ is the radius normalized by the halo's virial
radius. Using the Taylor series expansion for $\ln(1+x)/x$ we thus see
that the central potential of an NFW halo is given by
\begin{equation}
\Phi_\rmc \equiv \Phi(r=0) = -\left({V_{\rm max} \over 0.465}\right)^2\,.
\end{equation}
Hence, the maximum circular velocity of an NFW halo is a {\it direct}
measure of its central potential depth.

Throughout we use subscripts `0' to refer to the value at redshift
$z_0$, which we normally take to be the present day (i.e., $z_0=0$),
unless stated otherwise. We use time, $t$ , and redshift, $z$,
interchangeably as our `time coordinate', and use $t_0 - t$ to indicate
lookback time. Finally, given an ensemble $X = \{X_1,X_2,...X_n\}$, we
use $\langle X \rangle$ to indicate the ensemble's {\it average},
while $\langle X \rangle_{\rm med}$ is used to refer to its {\it
  median}. Typically we will plot medians whenever we also display
the halo-to-halo variance, and use averages when that is not the case.

\subsection{Semi-Analytical Model}
\label{sec:SAM}

One of the main goals of this paper is to present a universal model
for the MAHs and PWGHs of dark matter haloes as a function of halo
mass and cosmology. We trace the assembly of dark matter haloes using
a semi-analytical model based on merger trees constructed using the
extended Press-Schechter (EPS) formalism (Bond \etal 1991), which
provides the progenitor mass function (hereafter PMF), $n_{\rm
  EPS}(M_\rmp, z_1|M_0,z_0) \, \rmd M_\rmp$, that describes the
average number of progenitors of mass $M_\rmp \pm \rmd M_\rmp/2$ that
a descendant halo of mass $M_0$ at redshift $z_0$ has at redshift $z_1
> z_0$. Starting from some target host halo mass $M_0$ at $z_0$, one
can use this PMF to draw a set of progenitor masses ${M_{\rmp,1},
  M_{\rmp,2},...,M_{\rmp,N}}$ at some earlier time $z_1 = z_0 + \Delta
z$, where $\sum_{i=1}^{N} M_{\rmp,i} = M_0$ in order to assure mass
conservation. The time-step $\Delta z$ used sets the `temporal
resolution' of the merger tree, and may vary along the tree. This
procedure is then repeated for each progenitor with mass $M_{\rmp,i} >
M_{\rm res}$, thus advancing `upwards' along the tree. The minimum
mass $M_{\rm res}$ sets the `mass resolution' of the merger tree and
is typically expressed as a fraction of the final host mass $M_0$.

We construct our merger trees using the method of Parkinson, Helly \& Cole
(2008; hereafter P08), which uses the `binary method with accretion'
of Cole \etal (2000) combined with a PMF that is tuned to match
results from the Millennium Simulation (Springel et al. 2005), rather
than the PMF that follows from EPS. In particular, progenitor halo
masses are drawn from
\begin{equation}\label{nPMF}
n(M_\rmp,z_1|M_0,z_0) = n_{\rm EPS}(M_\rmp,z_1|M_0,z_0) \, G(M_\rmp,M_0,z_0)\,,
\end{equation}
where $G(M,M_0,z_0)$ is a perturbing function that is calibrated using
merger trees extracted from the Millennium simulation by Cole \etal
(2008), and which is given by
\begin{equation}\label{Gfunc}
G(M,M_0,z_0) = 0.57 \, \left[{\sigma^2(M) \over \sigma^2(M_0)}\right]^{0.19} \, 
\left[{\delta_\rmc(z_0) \over \sigma(M_0)}\right]^{-0.01}\,.
\end{equation}
Here $\delta_\rmc(z) = 1.686/D(z)$ is the initial overdensity required
for spherical collapse at redshift $z$, extrapolated to the present
time using linear theory, $\sigma^2(M)$ is the mass
variance\footnote{i.e., the variance in the linear fluctuation field
  when smoothed with a top-hat filter of size $R = (3M/4
  \pi\bar{\rho})^{1/3}$ with $\bar{\rho}$ the comoving density of the
  background.}, and $D(z)$ is the linear growth rate.  As shown in
Jiang \& van den Bosch (2014a), the P08 method yields halo merger
rates, mass accretion histories, and unevolved subhalo mass functions
(i.e., the mass function of subhaloes at accretion) that are all in
good agreement with numerical simulations. In addition, van den Bosch
\& Jiang (2014) have shown that it can also be used to predict
accurate {\it evolved} subhalo mass functions. Most importantly, even
though $G(M,M_0,z_0)$ was calibrated for the cosmology used to run the
Millennium simulation, the aforementioned studies have shown that it
yields equally accurate results for other $\Lambda$CDM cosmologies.
\begin{figure}
\centerline{\psfig{figure=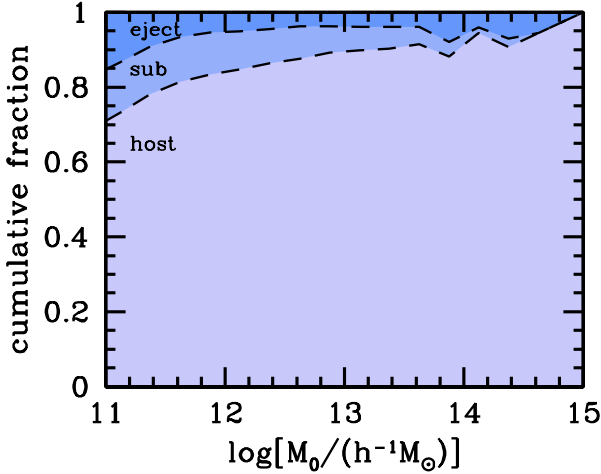,width=0.94\hssize}}
\caption{Cumulative fractions of host haloes, subhaloes and ejected
  haloes as function of halo mass in the Bolshoi simulation.}
\label{fig:frac}
\end{figure}

Throughout we always adopt a mass resolution of $M_{\rm res}/M_0 =
10^{-5}$ unless mentioned otherwise, and construct the merger trees
using the time stepping advocated in Appendix~A of P08 (which roughly
corresponds to $\Delta z\sim10^{-3}$; somewhat finer/coarser at
high/low redshift). In order to speed up the code, and to reduce
memory requirements, we down-sample the time resolution of each merger
tree by registering progenitor haloes every time step $\Delta t = 0.1
t_{\rm ff}(z)$\footnote{For the Bolshoi cosmology considered here,
  this results in $N=329$ time steps between $z=50$ and $z=0$.}. Here
$t_{\rm ff}(z) \propto (1+z)^{-3/2}$ is the free-fall time for a halo
with an overdensity of 200 at redshift $z$. Since the dynamical time
of a halo is of order the free-fall time, there is little added value
in resolving merger trees at higher time resolution than this. We have
verified that indeed our results do not change if we register our
merger trees using smaller time steps.

For a given merger tree, we determine the MAH by starting from the
final host halo at $z=z_0$, and tracing the tree back in time,
registering the mass of its main progenitor as function of
redshift. Next we use this $M(z)$ to compute $\Vmax(z)$ of the main
progenitor as function of redshift, using the method of Jiang \& van
den Bosch (2014b). In brief, we assume that dark matter haloes follow
NFW density profiles, for which $V_{\rm max}$ depends on the halo
virial velocity and halo concentration as given by
Eq.~(\ref{VmaxHost}). It is well known that the concentration of a
dark matter halo is strongly correlated with its MAH, in the sense
that haloes that assemble earlier are more concentrated (e.g., Navarro
\etal 1997; Wechsler \etal 2002; Ludlow \etal 2013). We use the model
of Zhao \etal (2009), according to which halo concentrations are given
by
\begin{equation}\label{concZhao}
c(M,t) = c(t,t_{0.04}) 
= 4.0\, \left[1+\left( {t \over 3.75\,t_{0.04}} \right)^{8.4} \right]^{1/8}\,,
\end{equation}
(but see \S\ref{sec:conc} below).  Here $t_{0.04}$ is the proper time
at which the host halo's main progenitor gained 4 percent of its mass
$M$ at proper time $t$, which we extract from the MAH. As shown in
Jiang \& van den Bosch (2014b) and van den Bosch \& Jiang (2014),
combining this methodology with a simulation-based description for how
$\Vmax$ of a subhalo evolves as it experiences mass stripping, yields
subhalo velocity functions, $\rmd N/\rmd\ln(\Vrat)$, that are in
excellent agreement with simulation results. In what follows we refer
to this semi-analytical model for computing the MAHs and PWGHs of dark
matter haloes as `MergerTrees',
\begin{figure*}
\centerline{\psfig{figure=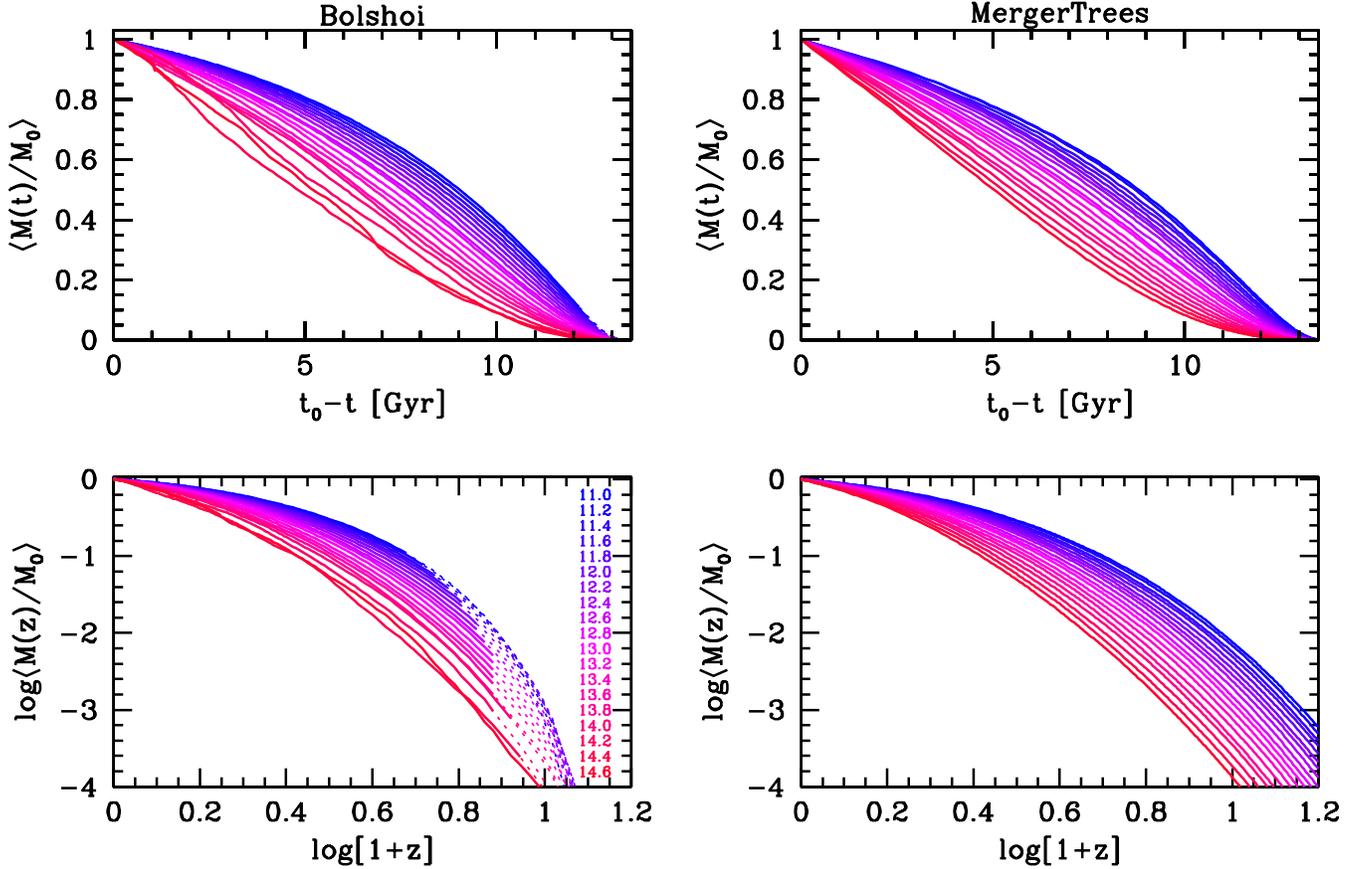,width=\hdsize}}
\caption{Average mass accretion histories for host haloes at
  $z=0$. Different colors indicate different host halo masses, with
  the corresponding value for $\log[M_0/(\Msunh)]$ indicated in the
  lower left-hand panel. Upper and lower panels plot $\langle M(t)/M_0
  \rangle$ as function of lookback time and $\log\langle M(z)/M_0
  \rangle$ as function of $\log[1+z]$, which accentuate the behaviors
  at late and early times, respectively. Panels on the left show the
  results obtained from the Bolshoi simulation, where each line is the
  average obtained from all haloes in a mass bin that is 0.2dex
  wide. Solid lines show the average MAHs over the range where the
  main progenitors of $>90\%$ of all host haloes can be traced. Dotted
  lines are the extensions obtained taking the average over all host
  haloes. Panels on the right show the results from MergerTrees, where
  each average is obtained using 2000 realizations.}
\label{fig:MAHs}
\end{figure*}

\subsection{Numerical Simulation}
\label{sec:SIM}

In order to test and, where needed calibrate, our semi-analytical
model we use the Bolshoi simulation (Klypin \etal 2011), which follows
the evolution of $2048^3$ dark matter particles using the Adaptive
Refinement Tree (ART) code (Kravtsov, Klypin \& Khokhlov 1997) in a
flat $\Lambda$CDM model with parameters $\Omega_{\rm m,0} = 1 -
\Omega_{\Lambda,0} = 0.27$, $\Omega_{\rm b,0} = 0.0469$, $h = H_0/(100
\kmsmpc) = 0.7$, $\sigma_8 = 0.82$ and $n_\rms = 0.95$ (hereafter
`Bolshoi cosmology').  The box size of the Bolshoi simulation is
$L_{\rm box} = 250 h^{-1} \Mpc$, resulting in a particle mass of
$m_\rmp = 1.35 \times 10^8 \Msunh$.

We use the publicly available halo catalogs and merger
trees\footnote{http://hipacc.ucsc.edu/Bolshoi/MergerTrees.html}
obtained using the phase-space halo finder \Rockstar (Behroozi \etal
2013a,b), which uses adaptive, hierarchical refinement of
friends-of-friends groups in six phase-space dimensions and one time
dimension. As demonstrated in Knebe \etal (2011, 2013), this results
in a very robust tracking of (sub)haloes (see also van den Bosch \&
Jiang 2014). In line with the halo definition used throughout this
paper, the \Rockstar haloes are defined as spheres with an average
density equal to $\Delta_{\rm vir} \rho_{\rm crit}$.  Details
regarding the construction of the merger trees can be found in
Behroozi \etal (2013b).

Throughout this paper we restrict ourselves to haloes that at $z=0$
have accumulated a mass $M_0 \geq 10^{11} \Msunh$ (corresponding to
$\geq 740$ particles per halo). Using the merger trees, we split the
population of $z=0$ haloes in three categories: 
\begin{itemize}
\item host haloes; these are distinct haloes that are not, and never
  have been, located within the virial radius of another, more massive
  halo.
\item subhaloes; these are haloes that at $z=0$ are located within the
  virial radius of another, more massive halo.
\item ejected haloes; these are haloes that at $z=0$ are distinct, but
  whose main progenitor has at one or more occasions passed through
  the virial region of a more massive halo. Ejected haloes are also
  sometimes called `backsplash' haloes.
\end{itemize}
Fig.~\ref{fig:frac} shows the cumulative fractions of these different
categories as function of their mass at $z=0$; host haloes clearly
dominate, with a fraction that increases from $\sim 70$ percent for
haloes with $M_0 = 10^{11} \Msunh$ to 100 percent at the massive end.
The remainder is split roughly equally between subhaloes and ejected
haloes. These statistics are in good agreement with previous studies
(e.g., Wang \etal 2009a).

For each host halo at $z=0$ we use the merger trees to determine their
MAH, $M(z)/M_0$, as well as their PWGH, $\Vmax(z)/V_{{\rm vir},0}$.
It is important to point out that in simulations the definition of
`main progenitor' is not without ambiguity. Whereas some studies
simply define it as the descendant's most massive progenitor (e.g.,
van den Bosch 2002; McBride \etal 2009; Fakhouri \etal 2010; Behroozi
\etal 2013c; Behroozi \& Silk 2014), others define it as the
progenitor that contributes most mass to the descendant (e.g.,
Wechsler \etal 2002; Zhao \etal 2003a,b, 2009; Giocoli \etal 2012). In
the EPS merger trees these two definitions are identical, but in
numerical simulations this is not necessarily the case. For example,
consider two progenitors of a descendant halo of mass $M$; progenitor
$A$ with mass $M_A = 0.53M$ and progenitor $B$ with mass
$M_B=0.51M$. Suppose that $B$ contributes its entire mass to its
descendant, whereas $A$ only contributes $0.49M$ (the remaining
$0.04M$ ending up just outside the boundary of the descendant
halo). In this case, $A$ is the most massive progenitor, whereas $B$
is the one that contributes most of its mass.  In this paper, we use
the publicly available merger trees from the Bolshoi simulation, and
always define the main progenitor as the most massive one. Although we
consider this the more preferable choice when comparing to EPS and
when using MAHs in the framework of galaxy formation, we acknowledge
that this somewhat ambiguous and that one can probably make equally
strong arguments in favor of picking the most-contributing progenitor
as the main progenitor. Although it is relatively rare that the
most-massive progenitor is different from the progenitor that provides
most mass, the different definitions for main progenitor result in
(average) MAHs that can be significantly different (see
\S\ref{sec:comparison}).

As demonstrated in Appendix~\ref{App:eject}, the MAHs of subhaloes and
ejected haloes are very different from those of host haloes. Since
this paper focuses on the MAHs and PWGHs of host haloes, and since
several studies have argued that galaxies that reside in ejected
haloes have properties that are more reminiscent of satellite galaxies
(those residing in subhaloes) than of central galaxies (e.g., Wang
\etal 2009b; Geha \etal 2012; Wetzel \etal 2014), we remove both
subhaloes and ejected haloes from our sample.
\begin{figure*}
\centerline{\psfig{figure=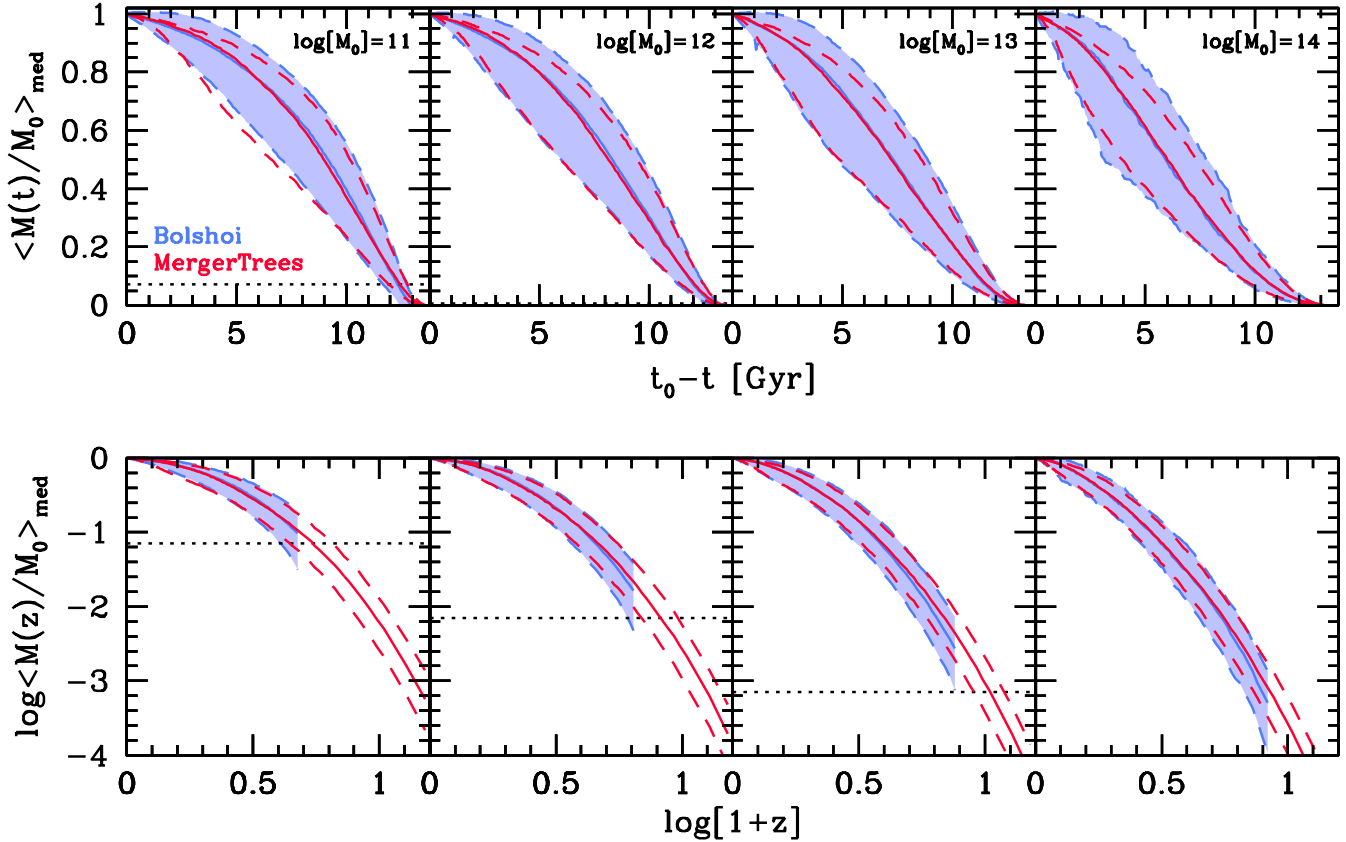,width=\hdsize}}
\caption{Median mass accretion histories. Comparison of Bolshoi
  results (blue) with results obtained using MergerTrees (red).
  Results are shown for four different mass bins, as indicated in the
  upper left corner of the upper panels. Solid lines depict the
  medians while the dashed lines indicate the corresponding 16 and 84
  percentiles (shaded in the case of Bolshoi). Note that the solid
  blue line is mostly invisible because the solid red line lies on top
  of it, indicating excellent agreement between simulations and
  MergerTrees.  Horizontal dotted lines (in black) indicate where the
  mass of the main progenitor drops below $7\times 10^9\Msunh$, which
  corresponds to 50 particles in the Bolshoi simulation, and roughly
  reflects the mass scale below which the simulation results are
  affected by resolution effects.}
\label{fig:compMAH}
\end{figure*}

\subsection{Why use a Semi-Analytical Model?}
\label{sec:motivation}

In an era in which numerical simulations have become common-place in
astrophysical research, one may wonder why one would resort to
semi-analytical techniques to study MAHs and/or PWGHs. In fact, there
are a number of reasons why we believe this to be important, useful
and even necessary.

The necessity comes mainly from the limited mass resolution in
numerical simulations: suppose we want to trace the main progenitor of
a halo back in time to when its $V_{\rm max}$ was roughly 10
percent of the present day value. For a Milky-Way sized halo this
implies resolving the main progenitor to when it roughly has a virial
temperature of $10^4$K. As we will see below, this requires resolving
the main progenitor of a $z=0$ halo of mass $M$ down to the point
where it drops below $10^{-4} M$. A reliable measurement of $V_{\rm
  max}$ of a simulated dark matter haloes requires that it is resolved
with at least 100 particles, which therefore implies a particle mass
$m_\rmp < 10^{-6} M$. Obtaining reliable statistics requires that the
simulation volume contains at least of order 1000 such host haloes,
which in turn implies a simulation volume $V > 1000/n(M)$, where $n(M)
= {\rm d}N/{\rm d}\ln M$ is the halo mass function. Using that the
total number of particles in the simulation box is $N_\rmp =
\Omega_\rmm \, \rho_{\rm crit} \, V/m_\rmp$, we then find that we need
\begin{equation}\label{Npart}
N_\rmp > 2.6 \times 10^{10} \, \left({\Omega_\rmm \over 0.3}\right) \,
\left[{ M\,n(M) \over 10^{9.5} h^{2}M_{\odot}{\rm Mpc}^{-3}}\right]^{-1}
\end{equation}
where we have used that in a $\Lambda$CDM cosmology, to good
approximation, $M\,n(M) \sim 10^{9.5} h^{2}M_{\odot}{\rm Mpc}^{-3}$
for present-day haloes with mass $M$ below the characteristic mass
$M^{*}$. For more massive haloes, the required number of particles
increases exponentially. Hence, proper statistics of MAHs (and PWGHs)
that are well resolved down to $10^{-4}M$ requires simulations that
are roughly three times the size of Millennium or Bolshoi, which are
among the largest simulations that have been run to date. Although it
is not infeasible that such simulations will be run in the not too
distant future, the computational costs will be enormous. This is
especially true in comparison to the analytical models, which can
construct thousands of merger trees in a matter of minutes or hours
(depending on the mass resolution used). Hence, it is clear that there
is great virtue in having a reliable, well calibrated semi-analytical
model, especially if it can be used for different cosmologies.

In addition, by resorting to simplified prescriptions of the underlying
dynamics, semi-analytical models are extremely useful for gaining
insight and understanding. Furthermore, although the accuracy of a
numerical simulation is only limited by its mass and force resolution,
there are non-trivial difficulties involved in identifying haloes and
in linking them to their earlier progenitors. As a result, depending
on the algorithms used, one can obtain merger trees that differ
substantially, even when applied to the same simulation (e.g. Helly
\etal 2003; Harker \etal 2006; Cole \etal 2008; Fakhouri \& Ma 2008;
Genel \etal 2008, 2009; Fakhouri, Ma \& Boylan-Kolchin 2010; Srisawat
\etal 2013).  This is a problem that will be difficult to overcome,
and implies that merger trees extracted from numerical simulations
have their own shortcomings and are not always completely
reliable. Hence, it is important and useful to have some alternatives
in the form of a semi-analytical model. 
\begin{figure}
\centerline{\psfig{figure=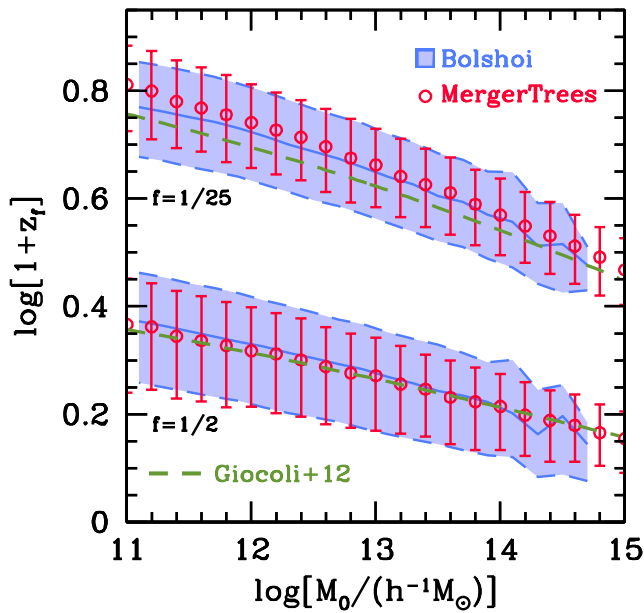,width=\hssize}}
\caption{Halo formation redshifts, $z_f$, as function of halo mass,
  $M_0$, for two values of $f$, as indicated. The solid blue line
  indicates the median in the Bolshoi simulation, with the shaded
  region, bounded by two dashed curves, indicating the 68 percent
  confidence intervals. Red, open circles are the median formation
  redshifts obtained with MergerTrees while the errorbars reflect the
  68 percent range. Green, dashed lines are the model predictions of
  Giocoli \etal (2012), and are shown for comparison. Note the
  excellent agreement between MergerTrees and simulation results.}
\label{fig:zformH}
\end{figure}
%


\section{Results}
\label{sec:res}

\subsection{Mass Accretion Histories}
\label{sec:MAH}

Fig.~\ref{fig:MAHs} plots the {\it average} MAHs for haloes of
different masses (different colors). We plot both $\langle M(t)/M_0
\rangle$ as function of lookback time (upper panels), as well as
$\log\langle M(z)/M_0 \rangle$ as function of $\log[1+z]$ (lower
panels). These accentuate the behavior of the MAHs at late and early
times, respectively. Panels on the left show the results obtained from
the Bolshoi simulation, where each line is the average obtained from
all haloes in a mass bin that is 0.2dex wide. Because of the limited
mass resolution, the MAHs become incomplete at early-times, when the
main progenitor mass drops below the mass resolution limit.  Following
Zhao \etal (2009), we therefore only plot (as solid lines) the average
MAHs up to the redshift or lookback time where the progenitors of
$>90\%$ of all host haloes in consideration can be traced (i.e., where
more than $90\%$ of all host haloes have a main progenitor that is
still resolved in the Bolshoi simulation). Dotted lines are the
extensions one obtains when taking the average over all host haloes,
assuming a main progenitor mass of zero whenever the MAH drops below
the resolution limit. The number of haloes used in the ensemble
averages ranges from $145126$ for the mass bin with
$\log[M_0/(\Msunh)] \in [10.9,11.1]$ to a meager 22 for the
$[14.5,14.7]$-bin, which explains the increasing `jaggedness' of the
average MAHs with increasing halo mass. The panels on the right show
the results from our semi-analytical model `MergerTrees', where each
curve is obtained by taking the average of 2000 MAHs for a host halo
with a mass equal to the midpoint of the logarithmic mass
bin\footnote{We have verified that drawing the 2000 halo masses from
  the full mass bin as sampled by the halo mass function yields
  results that are indistinguishable.}  ; i.e., for the bin
$\log[M_0/(\Msunh)] \in [10.9,11.1]$ this implies a halo mass of $M_0
= 10^{11}\Msunh$.

A few trends are apparent. First of all, it is clear that more massive
haloes assemble their mass later, consistent with hierarchical
structure formation and with numerous previous findings (e.g., vdB02;
Maulbetsch \etal 2007; Zhao \etal 2009; Fakhouri \etal 2010; Yang
\etal 2011; Giocoli \etal 2012). Secondly, overall the simulation and
MergerTrees results are in good, qualitative agreement, at least over
the range where $>90\%$ of the MAHs are resolved (i.e., are
represented by solid, rather than dotted lines).
Fig.~\ref{fig:compMAH} shows a more direct comparison, this time of
the {\it median} MAHs in the Bolshoi simulation (blue) and those
obtained using MergerTrees (red). Solid and dashed lines indicate the
median and the 68 percentile intervals, while the horizontal dotted
lines indicate the mass scale where $M(z)$ drops below $7\times
10^9\Msunh$. This corresponds to 50 particles per halo, and roughly
reflects the mass scale below which the Bolshoi simulation results
become strongly affected by resolution effects. Overall, the median
results obtained from MergerTrees are in excellent agreement with the
simulation results. Even the 68 percentile intervals are in good
agreement, although the simulation results reveal a more pronounced
tail towards higher $M(t)/M_0$ at late times (i.e., for lookback times
$t \lta 5$Gyr). This discrepancy is larger for more massive host
haloes and likely reflects subtle issues related to ejected haloes. For
example, consider a halo of mass $M_1$ that at time $t_1$ accreted
another halo of mass $M_2 \lta M_1$. If $M_2$'s orbital trajectory
places it outside of the virial radius of $M_1$ at the present day
then, in the simulation, halo $M_2$ is considered an ejected halo (and
thus removed from the sample), while $M_1$ is found to not have grown
in mass since time $t_1$. In our semi-analytical model, however, no
(sub)halo is ever ejected; hence, $M_1$ is considered to have grown in
mass from $M_1$ to $M_1 + M_2$ since time $t_1$. Hence, MergerTrees
will have a smaller fraction of haloes that have experienced little
recent growth than what is seen in simulations.

As a final comparison, Fig.~\ref{fig:zformH} plots different halo
assembly redshifts, $z_f$, defined as the redshift at which the main
progenitor of a halo of mass $M_0$ at $z=0$ first reaches a mass
$f\,M_0$. Results are shown for $f=0.5$ and $f=0.04$, as
indicated. The agreement between simulations (blue) and MergerTrees
(red) is excellent, both in terms of the medians and the 68 percentile
intervals. For comparison, the green, solid lines are the model
prediction of Giocoli \etal (2012): these are in excellent agreement
for $f=0.5$, but somewhat underpredict the $z_{0.04}$ compared to the
results from both MergerTrees and the Bolshoi simulation. Most likely
this is a manifestation of the fact that Giocoli \etal (2012) define
the main progenitor as the most-contributing progenitor, whereas we
define it to be the most-massive one (see \S\ref{sec:SIM}).
\begin{figure*}
\centerline{\psfig{figure=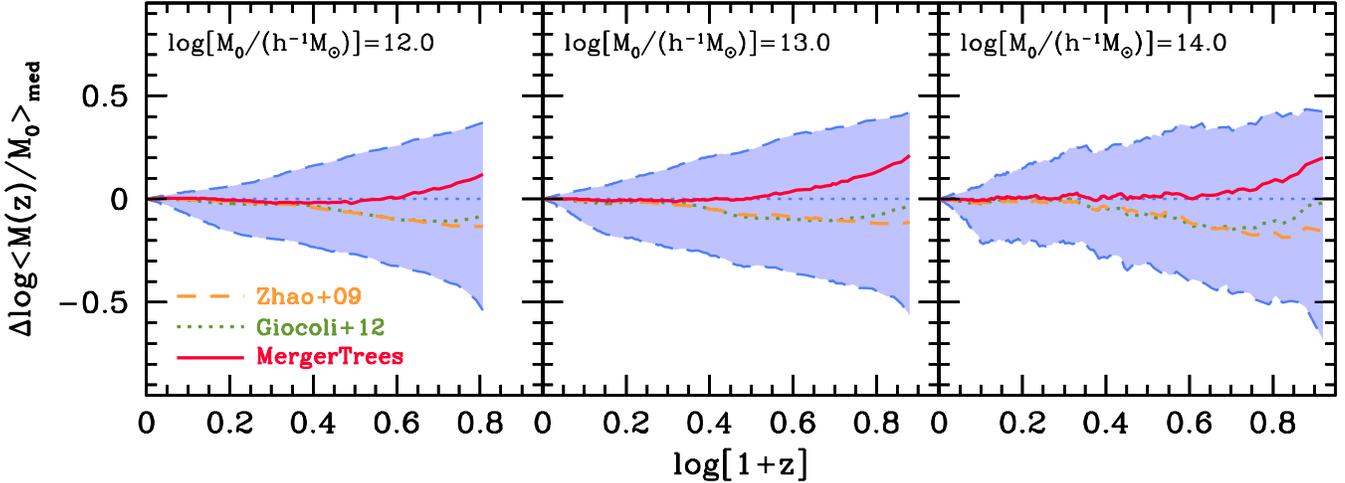,width=\hdsize}}
\caption{The differences $\Delta\log\langle M(z)/M_0\rangle_{\rm med}$
  between the median MAH obtained from the Bolshoi simulation and
  MergerTrees (red, solid lines), the universal model of Zhao \etal
  (2009; orange, dashed lines), and the universal model of Giocoli
  \etal (2012; green, dotted lines).  Results are shown for three halo
  masses as indicated, while the shaded region indicates the 68
  percentile interval. See text for discussion.}
\label{fig:comp_models}
\end{figure*}

\subsection{Comparison with Previous Studies}
\label{sec:comparison}

As mentioned in \S\ref{sec:intro}, a number of studies have already
investigated the MAHs of CDM haloes. Here we briefly describe some 
of the most relevant studies, and contrast their findings with those
presented here.

First, it is important to distinguish between two different kinds of
studies. On the one hand there are numerous studies that presented
fitting functions for the average and/or median MAHs (or their
time-derivatives) based on one particular numerical simulation (e.g.,
Neistein \& Dekel 2008a; McBride \etal 2009; Genel \etal 2009;
Fakhouri \etal 2010; Wu \etal 2013). These are only of limited use, as
they are only valid for the particular cosmology adopted in that
simulation. Even though most cosmological parameters are fairly well
constrained these days, the remaining uncertainties translate into
relatively large differences for the MAHs and PWGHs (see
Appendix~\ref{App:cosmo}). As a consequence, these studies can not be
used for a meaningful comparison with the results from the Bolshoi
simulation presented here.

In this section we therefore only focus on studies that have presented
a universal model for the MAHs of dark matter haloes as function of
halo mass, redshift and cosmology, and which can therefore be used to
predict the average or median MAHs of haloes in the Bolshoi
cosmology. The first such study was that by vdB02, who used EPS merger
trees to investigate how the {\it average} MAH scales with halo mass
and cosmology.  However, those merger trees were constructed using the
method developed by Somerville \& Kolatt (1999), which has since been
shown to yield MAHs that are systematically biased with respect to
simulation results (e.g., Zhang, Ma \& Fakhouri 2008; Jiang \& van den
Bosch 2014a). Indeed, we find the `universal function' of vdB02 to be
a poor fit to the average MAHs in the Bolshoi simulation, and will not
consider it further. More recently, Zhao \etal (2009; hereafter Z09)
used some elements from EPS theory to develop a universal model for
the MAHs of dark matter haloes, which they calibrated using a large
suite of numerical simulations for different cosmologies. A similar
approach was taken by Giocoli \etal (2012; hereafter G12), who derived
a method for computing the median MAHs that yield results that are
very similar to Zhao \etal, but using a prescription that is easier to
implement (see also Giocoli \etal 2013, where the same method was
extended to non-standard, coupled dark energy
cosmologies). Unfortunately, both Z09 and G12 defined their main
progenitor as the most-contributing one, which has to be taken into
account when comparing their results to ours.

Fig.~\ref{fig:comp_models} compares the predictions of the Z09 and G12
models with the median MAHs in the Bolshoi simulation and with those
obtained using MergerTrees. All model predictions are made for the
Bolshoi cosmology. Plotted are the differences, $\Delta\log\langle
M(z)/M_0\rangle_{\rm med}$, with respect to the median MAH of haloes
in the Bolshoi simulation. The shaded region, bounded by the two blue,
dashed curves, indicates the 68 percent halo-to-halo variance in the
Bolshoi simulation, while the red, orange and green lines are the
model predictions from MergerTrees, Z09 and G12, respectively. They
all agree with each other and with the Bolshoi simulation results for
$z \lta 1.5$.  At higher redshifts, the model predictions of Z09 and
G12 both start to underpredict $\langle M(z)/M_0\rangle_{\rm med}$.
Most likely this is a manifestation of the different methods used to
define the main progenitor, which is expected to yield MAHs that
diverge with increasing redshift. Although the effect is small
compared to the halo-to-halo variance, it is clear that one cannot use
the universal models of Z09 and/or G12 to describe the median MAHs
obtained with MergerTrees or obtained from simulations when the main
progenitor is defined to be the most-massive one.  In \S\ref{sec:uni}
we therefore present a new, universal model that is easy to use, and
that is valid for the main-progenitor definition adopted here.

Finally, Fig.~\ref{fig:comp_models} demonstrates that MergerTrees
yields MAHs that are in good agreement with the Bolshoi simulation
results out to $z \sim 3$, but then start to overpredict $\langle
M(z)/M_0\rangle_{\rm med}$. Note, though, that this discrepancy sets
in close to the mass resolution limit of the simulation (cf. dotted
curves in lower panels of Fig.~\ref{fig:compMAH}).  Hence, it remains
to be seen whether it reflects a true shortcoming of MergerTrees or
whether it is merely a manifestation of limited mass resolution in the
Bolshoi simulation.
\begin{figure*}
\centerline{\psfig{figure=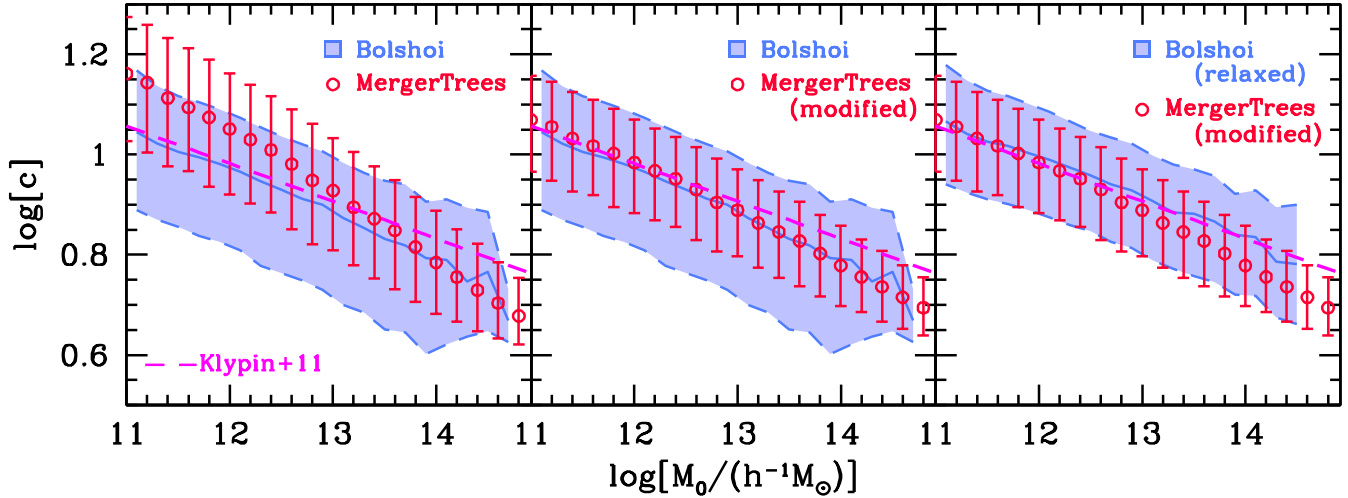,width=\hdsize}}
\caption{Concentration mass relations. The solid and dashed blue
  curves indicate the median and the 16 and 84 percentiles of the halo
  concentrations in the Bolshoi simulation. Red, open circles indicate
  the medians obtained using MergerTrees, with the errorbars
  reflecting the 68 percent interval. The dashed, magenta line is the
  median concentration-mass relation obtained from the Bolshoi
  simulation by Klypin \etal (2011), and is shown for comparison.  In
  the left-hand panel, all host haloes in the Bolshoi simulation have
  been used, and concentrations in MergerTrees have been obtained
  using Eq.~(\ref{concZhao}). In the middle panel, the Bolshoi data is
  the same, but this time concentrations in MergerTrees are computed
  using Eq.~(\ref{concZhaomod}). Finally, in the right-hand panel only
  the relaxed haloes in the Bolshoi simulation are used (see text for
  definition), while the MergerTrees results are the same as in the
  middle panel.}
\label{fig:conc}
\end{figure*}

\subsection{Halo Concentrations}
\label{sec:conc}

The main goal of this paper is to study the potential well growth
histories, $V_{\rm max}(t)/V_{\rm vir,0}$, of dark matter haloes.
Under the assumption that dark matter haloes have NFW density
profiles, the maximum circular velocity, $\Vmax$, is uniquely
specified by the halo's mass, $M$, and concentration, $c$ (see
\S\ref{sec:def}). As demonstrated in \S\ref{sec:MAH} above,
MergerTrees accurately describes the halo mass assembly history.
Hence, it is to be expected that it will also yield accurate PWGHs as
long as it can accurately predict halo concentrations.

In our model, halo concentrations are computed using the model of Zhao
\etal (2009), given by Eq.~(\ref{concZhao}), which links halo
concentration to $t_{0.04}$, the cosmic time at which the halo's main
progenitor first reaches a mass $0.04 M_0$. As shown in
Fig.~\ref{fig:zformH}, the $t_{0.04}$ obtained from MergerTrees are in
excellent agreement with those obtained from the Bolshoi simulation.
Hence, if the Z09 model is accurate, the halo concentrations predicted
by MergerTrees should also be in good agreement with those obtained
from the Bolshoi simulation.  

The solid, blue line in Fig.~\ref{fig:conc} shows the {\it median}
concentration-mass relation for $z=0$ dark matter haloes in the
Bolshoi simulation, while the blue shaded region indicates the 68
percentile interval. These concentrations have been obtained using the
method of Klypin \etal (2011), i.e., they are inferred from $V_{\rm
  max}$ and $M$ under the assumption that haloes follow an NFW
profile, without having to actually fit the halo's density
profile\footnote{We have compared these to the concentrations obtain
  by fitting the actual halo density profiles and find results that
  are almost indistinguishable.}.  For comparison, the magenta line
corresponds to
\begin{equation}\label{cKlypin}
c = 9.60 \, \left({M_0 \over 10^{12} \Msunh}\right)^{-0.075}
\end{equation}
which is the median $z=0$ concentration-mass relation obtained from
the same Bolshoi simulation by Klypin \etal (2011). Although both
results agree at the low mass end, the concentration-mass relation of
Klypin \etal is somewhat shallower than that inferred here.  We
emphasize, though, that Klypin \etal used the Bounded-Density-Maxima
(BDM) halo finder of Klypin \& Holzman (1997), whereas the results
presented here are based on \Rockstar. Since different halo finders
assign slightly different masses to identical haloes (e.g., Knebe
\etal 2011), this is most likely the cause of the small discrepancy
seen at the massive end.

The red, open circles with errorbars in the left-hand panel of
Fig.~\ref{fig:conc} are the median and 68 percentile intervals
obtained from MergerTrees, using 2000 Monte Carlo realizations per
halo mass bin. Although the median is in good agreement with the
simulation results for $M_0 \gta 5 \times 10^{13} \Msunh$, it is
systematically offset to larger concentrations for less massive
haloes. The origin of this discrepancy most likely results from the
fact that Zhao \etal (2009) calibrated their model using values for
$t_{0.04}$ that are obtained from MAHs in which the main progenitor is
defined to the most-contributing one, as opposed to the most-massive
one. As is evident from Fig.~\ref{fig:zformH} these different
definitions result in different values for $t_{0.04}$, and thus in
different predictions for the halo concentrations. Since it is prudent
for estimating reliable $V_{\rm max}$ that our halo concentrations are
accurate, we modify Eq.~(\ref{concZhao}) so that we reproduce the
median concentration-mass relation in the Bolshoi simulation.  The red
circles with errorbars in the middle panel of Fig.~\ref{fig:conc} show
the results obtained from MergerTrees if we assign halo concentrations
using
\begin{equation}\label{concZhaomod}
c(t,t_{0.04}) = 4.0\, 
\left[1+\left( {t \over 3.40\,t_{0.04}} \right)^{6.5} \right]^{1/8}.
\end{equation}
instead of Eq.~(\ref{concZhao}). As is apparent, this modified model
yields median halo concentrations that are in excellent agreement with
the Bolshoi results. In what follows we will adopt this modified model
throughout.

Although the median of the modified model is in excellent agreement
with the simulation results, it predicts significantly less scatter. In
particular, the simulations reveal concentration distributions at
fixed mass, $\calP(c|M)$, that have an extended tail towards lower
concentrations. Such a tail is not present in the PDFs obtained using
MergerTrees, which instead are close to log-normal.  Several studies
have shown that a low-concentration tail in $\calP(c|M)$ is due to
non-relaxed haloes, which are haloes that are temporarily out of
virial equilibrium due to merger activity, and therefore poorly
described by an NFW profile (e.g., Macc\`io \etal 2007, 2008; Neto
\etal 2007; Ludlow \etal 2013). To test this, the blue shaded
region in the right-hand panel indicates the 68 percentile interval
obtained using only {\it relaxed} haloes in the Bolshoi simulation.
Following Ludlow \etal (2013), these are defined as haloes for which
the ratio of kinetic to potential energy, $T/\vert U \vert < 0.0625$,
and the distance between halo barycenter and the location of the
potential minimum is less than 7 percent of the virial radius. Roughly
$84\%$ of the Bolshoi haloes in the mass range $10^{11} \Msunh \leq
M_0 \leq 10^{15} \Msunh$ meet these criteria, and their $\calP(c|M)$
is consistent with a log-normal with a scatter $\sigma_{\log c} \simeq
0.11$. Note that MergerTrees somewhat underpredicts the median
concentration of relaxed haloes at the massive end, and that it also
predicts that the amount of scatter decreases with increasing mass,
from $\sigma_{\log c} \simeq 0.10$ for $M_0 = 10^{11} \Msunh$ to
$0.05$ for $M_0 = 10^{15} \Msunh$. Although the latter is not apparent
in the Bolshoi simulations, it is in excellent agreement with Neto
\etal (2007), who, using the Millennium simulation, found that {\it
  relaxed haloes} have log-normal distributions of halo concentration
with a scatter that decreases from $\sigma_{\log c} = 0.11$ for
$M_0\sim 10^{12} \Msunh$ to $0.06$ for $10^{15}\Msunh$. The origin of
this discrepancy between the results of Neto \etal (2007) and that of
the Bolshoi simulation presented here is unclear.
\begin{figure*}
\centerline{\psfig{figure=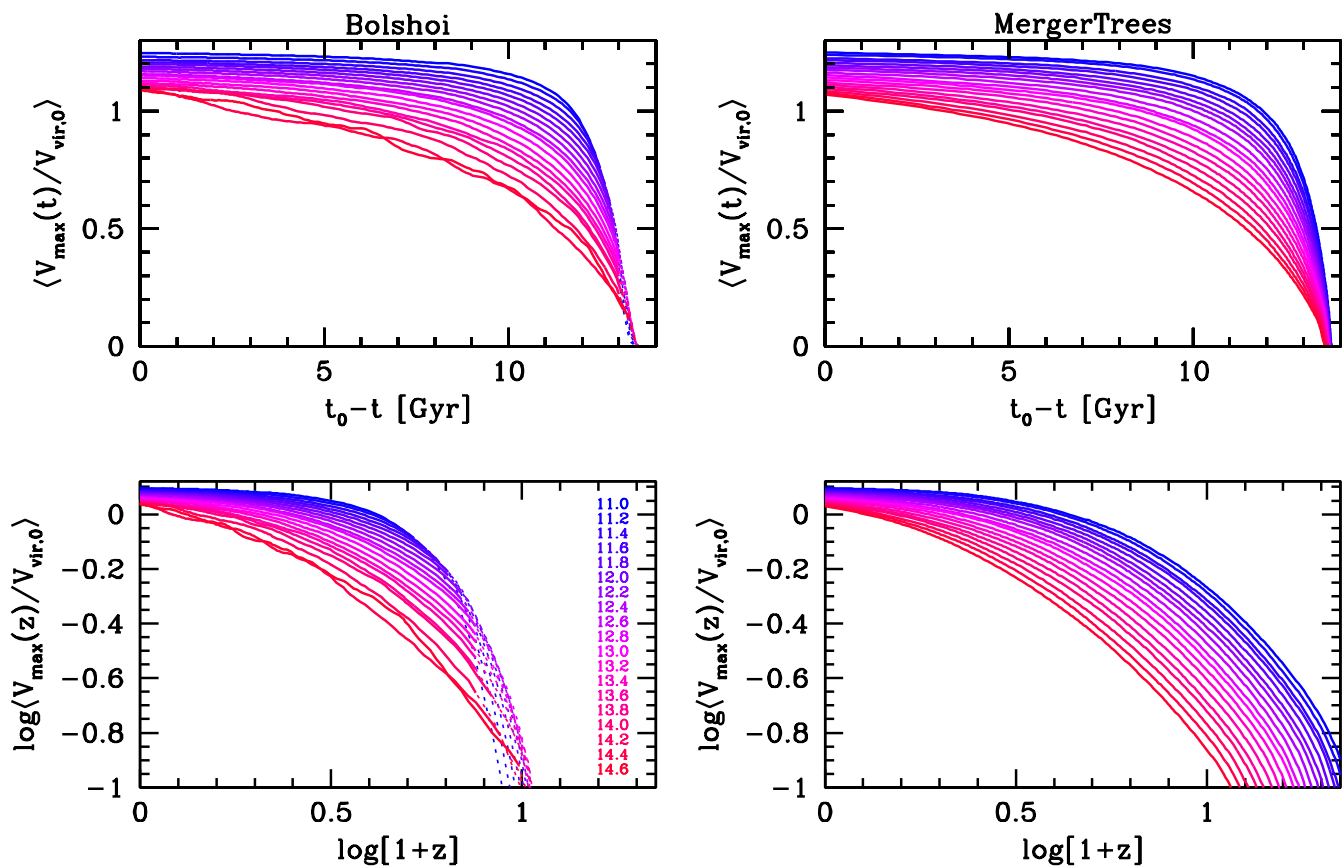,width=\hdsize}}
\caption{Same as Fig.~\ref{fig:MAHs}, but for the average PWGHs, $V_{\rm
    max}(t)/V_{{\rm vir},0}$.}
\label{fig:VAHs}
\end{figure*}

We conclude that our semi-analytical model with the modified
$c(t,t_{0.04})$ relation of Eq.~(\ref{concZhaomod}) yields a median
concentration-mass relation that is in good agreement with the Bolshoi
simulation, but with a scatter that is somewhat too small.
Consequently, our model will also somewhat underestimate the
halo-to-halo variance in PWGHs. An alternative would be to remove the
unrelaxed haloes from the sample of simulated haloes, which is the
approach that was taken by Ludlow \etal (2013). However, this would
remove the $\sim 15$ percent of the haloes that experienced most
growth in the recent past, and would therefore seriously bias the
median and average results. Since (recent) merging is an essential
ingredient of the CDM paradigm, we want our semi-analytical model to
be representative of the entire population of host haloes, not only of
the subset of relaxed host haloes. As we demonstrate in the next
section, our model accurately reproduces the median PWGHs of Bolshoi
host haloes (including the unrelaxed ones), and only marginally
underestimates the halo-to-halo variance.

\subsection{Potential Well Growth Histories}
\label{sec:PWGH}

Fig.~\ref{fig:VAHs} plots the {\it average} PWGHs for different bins in
host halo mass (different colors). Results are shown for both the
Bolshoi simulation (left-hand panels) and for MergerTrees (right-hand
panels).  Similar to Fig.~\ref{fig:MAHs}, upper and lower panels plot
the results linearly and logarithmically to better accentuate the
behavior at late and early times, respectively. As before, solid lines
indicate the results up to the redshift or lookback time where the
progenitors of $>90\%$ of all host haloes in consideration can be
traced, while dotted lines show the extensions obtained averaging over
all host haloes while assuming $V_{\rm max}=0$ whenever the MAH drops
below the resolution limit.  Qualitatively, the simulation and
MergerTrees results are in good agreement, at least over the range
where $>90\%$ of the MAHs are resolved (i.e., are represented by
solid, rather than dotted lines).

Note that, contrary to the MAHs shown in Fig.~\ref{fig:MAHs}, the PWGHs
do not all converge to unity at $z=0$. Rather, because of the
concentration-mass relation, the present-day ratio of $V_{\rm
  max}/V_{\rm vir}$ is larger for more concentrated (less massive)
haloes. Alternatively, we could have opted to define the PWGHs as
$V_{\rm max}(t)/V_{{\rm max},0}$, rather than $V_{\rm max}(t)/V_{{\rm
    vir},0}$.  However the former contains less information as it is
trivially recovered from the latter by multiplying by $V_{{\rm
    vir},0}/V_{{\rm max},0}$ (i.e., dividing by the PWGH at $z=0$.).

\begin{figure}
\centerline{\psfig{figure=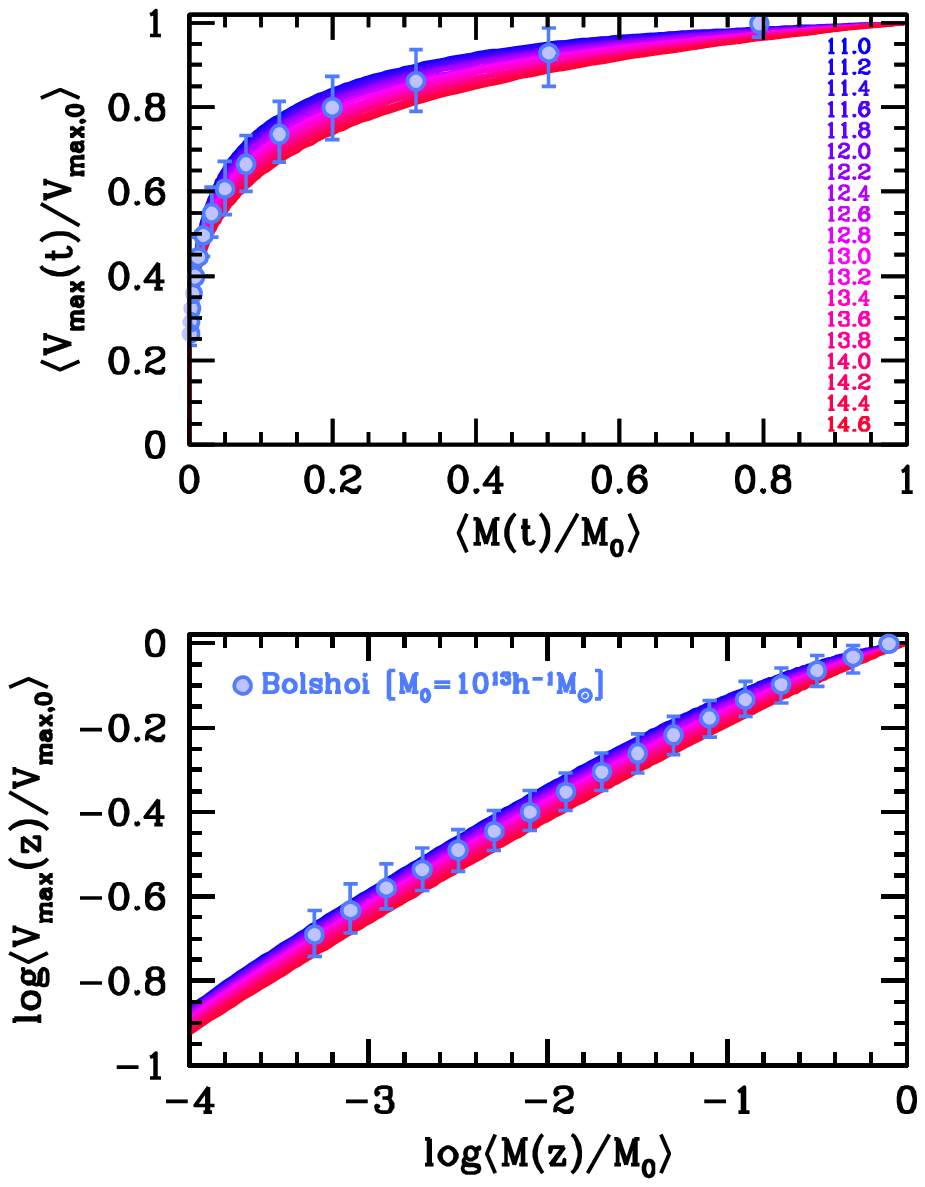,width=\hssize}}
\caption{The average $V_{\rm max}(t)/V_{{\rm max},0}$ as function of
  the average $M(t)/M_0$. Lines of different color, which can barely
  be discerned from one another, correspond to the results obtained
  using MergerTrees for haloes of different mass, $M_0$; color coding
  is the same as in Figs.~\ref{fig:MAHs} and~\ref{fig:VAHs}, and is
  indicated in the top panel. Solid circles are results obtained from
  the Bolshoi simulation for haloes with $\log[M_0/(\msunh)] \in
  [12.9,13.1]$; only plotting results for the range where $>90\%$ of
  the MAHs are resolved. Errorbars mark the 68 percent interval from
  the distribution of {\it individual} MAHs in this mass range. These
  results clearly illustrate the inside-out assembly of dark matter
  haloes, with the central potential well forming well before most of
  the mass is in place.}
\label{fig:MAHVAH}
\end{figure}

It is clear from Fig.~\ref{fig:VAHs} that less massive haloes
establish their central potential wells earlier, and that this
rank-order is preserved at all times; i.e., at any given time the main
progenitor of what ends up being a less massive halo at $z=0$ has
already build up a larger fraction of its final, central potential
well. Note that this rank-order-preservation is violated by the dotted
curves in the left-hand panels, which stresses the importance of only
averaging results over ensembles that are well resolved.  Upon
comparing Figs.~\ref{fig:MAHs} and~\ref{fig:VAHs}, it is evident that
dark matter haloes establish their potential wells before they have
accreted a major fraction of their mass (see also Li \etal 2007 and
Boylan-Kolchin \etal 2010). This behavior is especially evident from
Fig.~\ref{fig:MAHVAH}, which plots $\langle V_{\rm max}(t)/V_{{\rm
    max},0} \rangle$ as function of $\langle M(t)/M_0 \rangle$.
Colored lines are the averages obtained using MergerTrees for
different halo masses (as indicated), while the blue dots indicate the
averages obtained from the Bolshoi simulation for haloes with $12.9 <
\log[M_0/(\Msunh)] \leq 13.1$.  Errorbars indicate the 68 percent
confidence interval of the Bolshoi haloes, and indicate that the
halo-to-halo variance in this plot is small. Moreover, the MergerTrees
results show that there is only a weak dependence on halo mass, so
that the vast majority of all dark matter haloes will lie along the
colored band. The reason why this relation is so tight is easy to
understand from the fact that
\begin{eqnarray}\label{Vrat}
\lefteqn{ {V_{\rm max}(t) \over V_{{\rm max},0}} = 
{V_{\rm max}(t) \over V_{\rm vir}(t)} \, {V_{\rm vir}(t) \over V_{{\rm vir},0}} \, 
{V_{{\rm vir},0} \over V_{{\rm max},0}} } \nonumber \\
 & & = {f(c) \over f(c_0)} \, \left[{M(t) \over M_0}\right]^{1/3} \,
\left[{\Delta_{\rm vir}(t) \over \Delta_{{\rm vir},0}}\right]^{1/6} \,
\left[{H(t) \over H_0}\right]^{1/3} \,,
\end{eqnarray}
where $c$ and $c_0$ are the halo concentrations corresponding to
$M(t)$ and $M_0$, respectively, and we have used
Eqs.~(\ref{VmaxHost})--(\ref{Vvir}). This shows that, to first order,
$V_{\rm max}(t)/V_{{\rm max},0} \propto [M(t)/M_0]^{1/3}$.  Deviations
from this simple power-law reflect the concentration-mass-redshift
relation of dark matter haloes and the fact that the virial
overdensity and the Universe's expansion rate depend on redshift, but
these dependencies are relatively weak.  As a rule of thumb, the
maximum circular velocity of a halo's main progenitor is already half
the present day value by the time it has accreted only about 2 percent
of its final mass. In addition, when the halo has assembled half its
mass, its $V_{\rm max}$ is already at about 90 percent of its final
value. More accurate results can be obtained using the universal model
presented in \S\ref{sec:uni} below.

\begin{figure*}
\centerline{\psfig{figure=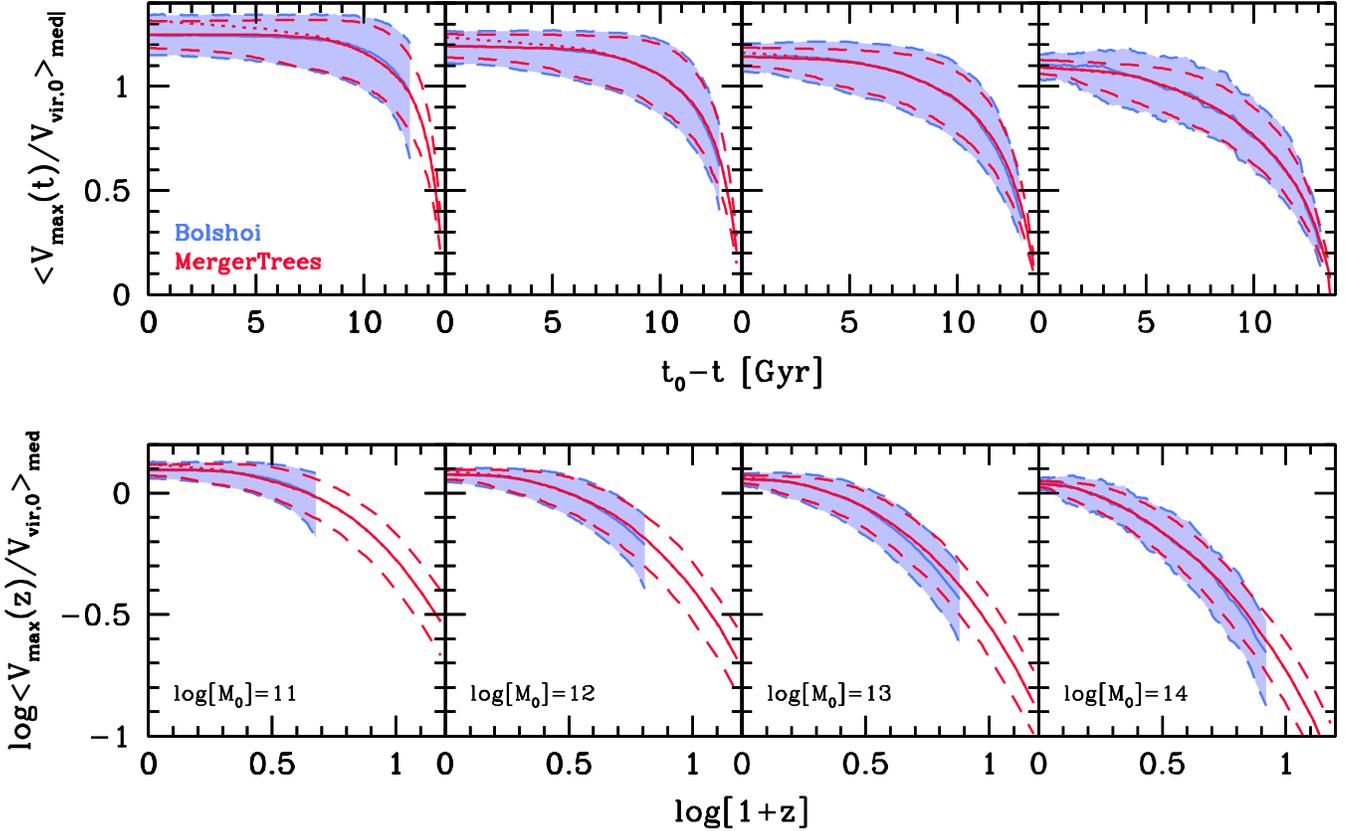,width=\hdsize}}
\caption{Same as Fig.~\ref{fig:compMAH}, but for the median PWGHs,
  $V_{\rm max}(t)/V_{{\rm vir},0}$. The dotted, red lines show the
  median PWGHs obtained from MergerTrees with the original
  concentration-mass-redshift relation of Zhao \etal (2009;
  Eq.~[\ref{concZhao}]), which somewhat overshoots $V_{\rm max}$ of
  low mass haloes at late times, simply because it overpredicts their
  concentrations (see \S\ref{sec:conc}).}
\label{fig:compVAH}
\end{figure*}

Fig.~\ref{fig:compVAH} shows a more direct comparison of the {\it
  median} PWGHs in the Bolshoi simulation (blue) and those
obtained using MergerTrees (red).  As in Fig.~\ref{fig:VAHs}, upper
and lower panels plot the linear and logarithmic PWGHs,
respectively. Solid and dashed lines indicate the median and the 68
percentile intervals. Except where the results drop below the mass
resolution limit, the median results obtained from MergerTrees are in
excellent agreement with the simulation results.  Note that
MergerTrees also predicts 68 percentile intervals that are in
extremely good agreement with Bolshoi, even though it underpredicts
the amount of scatter in the halo concentrations. Finally, the dotted,
red lines show the results obtained from MergerTrees with the original
concentration-mass-redshift relation of Zhao \etal (2009;
Eq.~[\ref{concZhao}]). As you can see, this somewhat overshoots
$V_{\rm max}$ of low mass haloes at late times, simply because it
overpredicts their concentrations (see \S\ref{sec:conc}).


\section{Universal Model}
\label{sec:uni}

Having demonstrated that our semi-analytical model can successfully
reproduce the MAHs and PWGHs in the Bolshoi simulation, we now use it
to develop a `universal model' that can be used to quickly compute the
average or median MAH and PWGH of a dark matter halo of any given mass
and for any (realistic) $\Lambda$CDM cosmology, without having to run
a numerical simulation or a set of halo merger trees.

The first step is to realize that once we have a model for the MAH, it
is straightforward to compute the corresponding PWGH using
\begin{eqnarray}\label{Vratb}
\lefteqn{{V_{\rm max}(t) \over V_{{\rm vir},0}} = {V_{\rm max}(t) \over V_{\rm vir}(t)} \,
 {V_{\rm vir}(t) \over V_{{\rm vir},0}} } \nonumber \\
 & & = 0.465 \, f(c) \, \left[{M(t) \over M_0}\right]^{1/3} \,
\left[{\Delta_{\rm vir}(t) \over \Delta_{{\rm vir},0}}\right]^{1/6} \,
\left[{H(t) \over H_0}\right]^{1/3}
\end{eqnarray}
where $c$ is the concentration of the main progenitor of mass $M$ at
time $t$ (cf. Eq.~[\ref{Vrat}]). Hence, the PWGH follows directly from
the MAH and a model for the concentration-mass-redshift relation.  In
fact, the halo concentration can be computed directly from the MAH
using Eq.~(\ref{concZhaomod}), so that all we really need is a model
for the MAH.

Unfortunately, as discussed in \S\ref{sec:comparison}, the `universal
models' for the MAHs of dark matter haloes developed by Zhao \etal
(2009) and Giocoli \etal (2012) are inadequate to describe our MAHs
because they used a different definition for the main progenitor. We
therefore start by developing a new, universal model, which then
serves as the basis for computing the PWGHs using Eq.~(\ref{Vratb})
above. Our model is motivated by the fact that simulations and EPS
studies have shown that the unevolved subhalo mass function is
(almost) universal (Lacey \& Cole 1993; Zentner \& Bullock 2003; van
den Bosch \etal 2005; Giocoli, Tormen \& van den Bosch 2008; Li \& Mo
2009; Yang \etal 2011). This means that, statistically speaking, all
haloes grow in mass by accreting the same haloes when expressed in
terms of their normalized mass $M_\rmp/M_0$ (here $M_\rmp$ is the
progenitor mass, and $M_0$ is the final host mass). Hence, average
MAHs for haloes of different $M_0$ and/or different cosmology only
differ from each other because they accrete those progenitors at
different times (see also Neistein \& Dekel 2008a; Genel \etal 2009;
Neistein \etal 2010; Yang \etal 2011). This suggests that one should
be able to transform the average or median MAH for a halo of mass,
$M_{0,\rmr}$, and cosmology, $\calC_\rmr$, to that of a halo of
another mass and cosmology, ($M_{0,\rmt},\calC_\rmt$), via a simple
transformation of the time-coordinate (cf., Neistein \etal 2010). Here
the subscripts `r' and `t' refer to the `reference' and `target' MAH,
respectively.  Using $\psi$ as shorthand for either the median or the
average MAH, i.e., $\psi(z) \equiv \langle M(z)/M_0 \rangle_{\rm med}$
or $\psi(z) \equiv \langle M(z)/M_0 \rangle$, respectively, one can
then write that
\begin{equation}
\psi(z|M_{0,\rmt},z_{0,\rmt},\calC_\rmt) = \psi(z'|M_{0,\rmr},z_{0,\rmr},\calC_\rmr)\,,
\end{equation}
and all that remains is to find the appropriate time-transformation,
$z = z(z')$, and to identify a reference MAH. 

\subsection{Self-similarity in time}
\label{sec:selfsim}

In the EPS formalism, there is a natural variable for which the
(conditional) mass function of dark mater haloes is invariant with
halo mass and cosmology. This variable is $\delta_\rmc(z)/\sigma(M)$,
where $\delta_\rmc(z) = 1.686/D(z)$ and $\sigma^2(M)$ is the mass
variance. This suggests that a natural choice for the time
transformation $z' \rightarrow z(z')$ is given by $\omega_\rmt(z) =
\omega_\rmr(z')$, where
\begin{equation}\label{omegaf}
\omega(z) = \omega(z|M,M_0,z_0) \equiv {\delta_\rmc(z) -
  \delta_\rmc(z_0) \over \sqrt{\sigma^2(M) - \sigma^2(M_0)}}\,.
\end{equation}
Indeed, as shown by Lacey \& Cole (1993), the distribution of halo
formation times takes on a form that is (almost) independent of halo
mass and/or cosmology\footnote{$\omega$ only has a weak dependence on
  the slope of the matter power spectrum.} when expressed in terms of
$\omega$.  Although we find that this scaling (also promoted by
Neistein \etal 2010) captures much of the trends that MAHs display
with mass and cosmology, it lacks sufficient precision and becomes
progressively worse for larger mass and/or cosmology differences
between reference and target, especially at larger redshifts. This
should not come entirely as a surprise. After all, it is well-known
that EPS predicts halo assembly to occur later than what is found in
numerical simulations (e.g. van den Bosch 2002a; Lin, Jing \& Lin
2003; Neistein \etal 2006). Related to this is the fact that EPS, when
based on spherical collapse, yields (conditional) halo mass functions
that fail to accurately match simulation data. In fact, this is the
reason why the Parkinson \etal (2008) algorithm that we use to
construct our merger trees relies on a progenitor mass function that
multiplies the EPS prediction with the perturbing function
$G(M,M_0,z_0)$ given by Eq.~(\ref{Gfunc}).
\begin{figure*}
\centerline{\psfig{figure=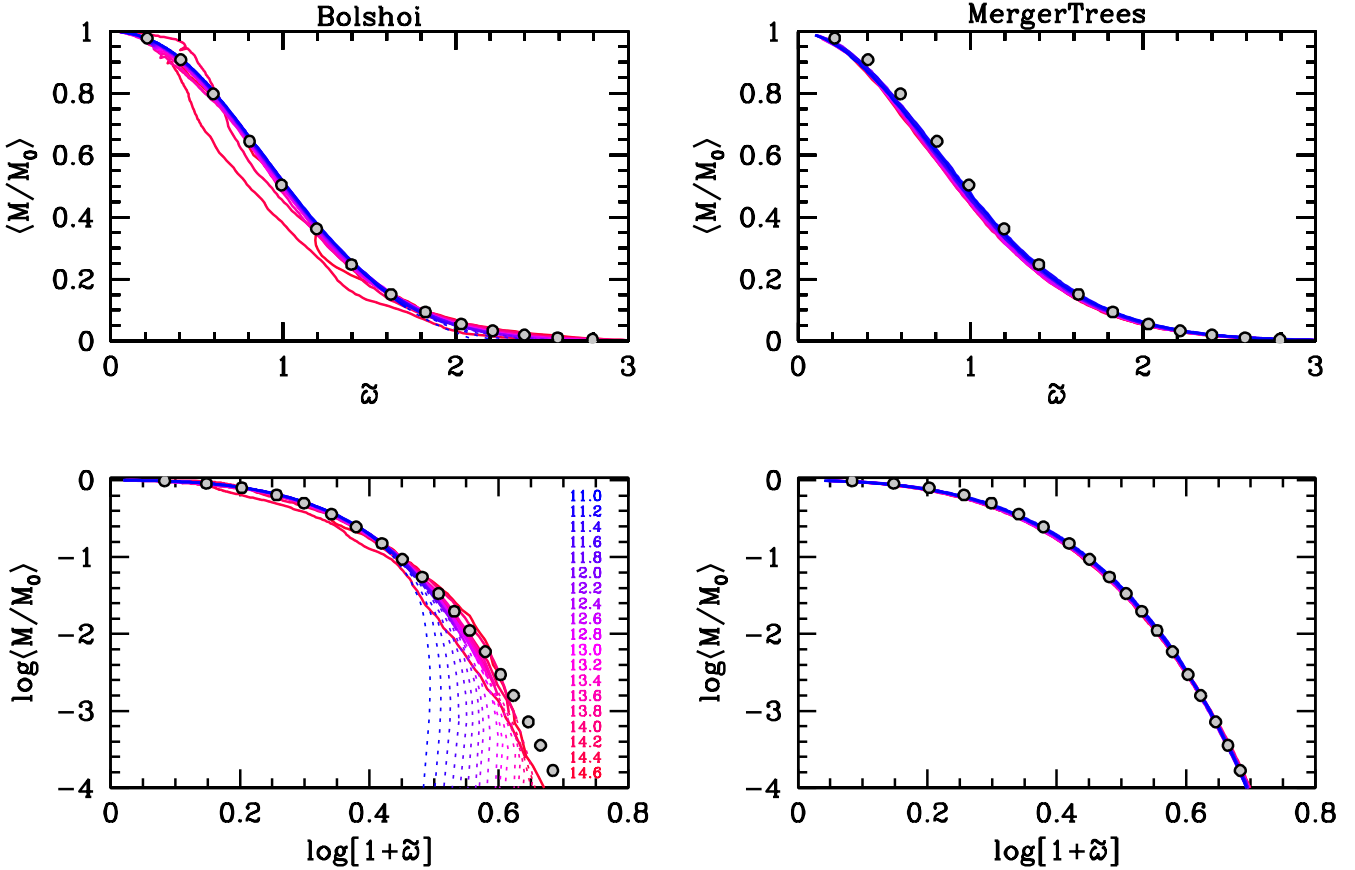,width=\hdsize}}
\caption{Same as Fig.~\ref{fig:MAHs}, but this time the average MAHs
  are plotted as function of the time variable $\tomega$, defined by
  Eq.~(\ref{omegafmod}). Note how all MAHs collapse on top of each
  other, indicating that the average (and median) MAHs of dark matter
  haloes have a universal shape. The few curves in the left-hand panel
  that do not fall on top of the others correspond to massive haloes,
  for which the averages are noisy due to small number statistics.
  The gray circles indicate the universal relation between $\tomega$
  and $\psi = \langle M/M_0\rangle$, which is well fitted by
  Eq.~(\ref{fitfunc}) and with the best-fit parameters listed in
  Table~1.}
\label{fig:wf}
\end{figure*}

Hence, a logical next step is to consider a time-transformation
$\tomega_\rmt(z) = \tomega_\rmr(z')$, where
\begin{equation}\label{omegafmod}
\tomega(z|M,M_0,z_0) \equiv \omega(z|M,M_0,z_0) \, G^{\gamma}(M,M_0,z_0)
\end{equation}
with $\gamma$ a free parameter. Indeed, after some trial and error we
find that using $\gamma=0.4$ results in extremely accurate
transformations from one MAH to another.  In particular, this
transformation not only works for the Bolshoi simulation, but also can
be used to transform MAHs from one $\Lambda$CDM cosmology to the
other. It implies that the average and/or median MAH has a universal
form when written as $\psi(\tomega)$.  Since $\tomega$ itself depends
on $\psi$ (i.e., $M = \psi \, M_0$), this universal form is a
parametric equation that has to be solved numerically (see
Appendix~\ref{App:model} for details).

Fig.~\ref{fig:wf} shows the same average MAHs as in
Fig.~\ref{fig:MAHs}, but this time plotted as function of
$\tomega$. Left and right-hand panels plot MAHs obtained from Bolshoi
and MergerTrees, respectively. As in Fig.~\ref{fig:MAHs}, dotted
curves are the extensions of the Bolshoi MAHs to the redshift range
where fewer than 90\% of the individual MAHs can be traced, and are
therefore heavily influenced by resolution effects.  Note how all MAHs
fall on top of each other, reflecting the universal character of
$\psi(\tomega)$.

In order to characterize the universal, parametric form of
$\psi(\tomega)$ we fit the inverse relation, $\tomega(\psi)$, using
the following functional form
\begin{equation}\label{fitfunc}
\calF(\psi) \equiv a_1 \, \left[1 - a_2 \log\psi\right]^{a_3} \, 
(1 - \psi^{a_4})^{a_5}
\end{equation}
where $(a_1,a_2,...,a_5)$ are treated as free parameters. Since the
Bolshoi results are arguably more reliable than MergerTrees, we fit
the free parameters using Bolshoi data, but only at $z \leq 2$
($\tomega \lta 1.5$). At higher redshifts, the Bolshoi data either
suffers from limited mass resolution (in the case of low mass haloes),
or from a small sample size (in the case of massive
haloes). Therefore, we complement the simulation data with MergerTrees
results for $z > 2$.  The resulting best-fit parameters are given in
Table~1, both for the average and the median MAHs. The corresponding
best-fit relation for the average MAH is shown as solid dots in
Fig.~\ref{fig:wf}, which accurately fit the Bolshoi and MergerTrees
results. 
\begin{table}\label{tab:param}
\caption{Parameters of Universal MAHs}
\begin{center}
\begin{tabular}{rccccc}
\hline\hline
 MAH    & $a_1$ & $a_2$ & $a_3$ & $a_4$ & $a_5$ \\
\hline
average & 3.295 & 0.198 & 0.754 & 0.090 & 0.442 \\
median  & 1.928 & 0.424 & 0.768 & 0.148 & 0.310 \\
\hline\hline
\end{tabular}
\end{center}
\medskip
\begin{minipage}{\hssize}
  Best-fit parameters for Eq.~(\ref{fitfunc}) describing the average
  (upper row) and median (lower low) MAHs.
\end{minipage}
\end{table}

As detailed in Appendix~\ref{App:model}, by numerically solving
\begin{equation}\label{unieq}
\calF(\psi) = \tomega(\psi,z)\,,
\end{equation}
either for $\psi$ at given $z$, or for $z$ at given $\psi$, one can
compute the average or median MAHs for a host halo of any mass, $M_0$,
at any redshift, $z_0$, and for any ($\Lambda$CDM) cosmology. Using
Eq.~(\ref{Vratb}), these can subsequently be used to compute the
corresponding average or median PWGH.
\begin{figure*}
\centerline{\psfig{figure=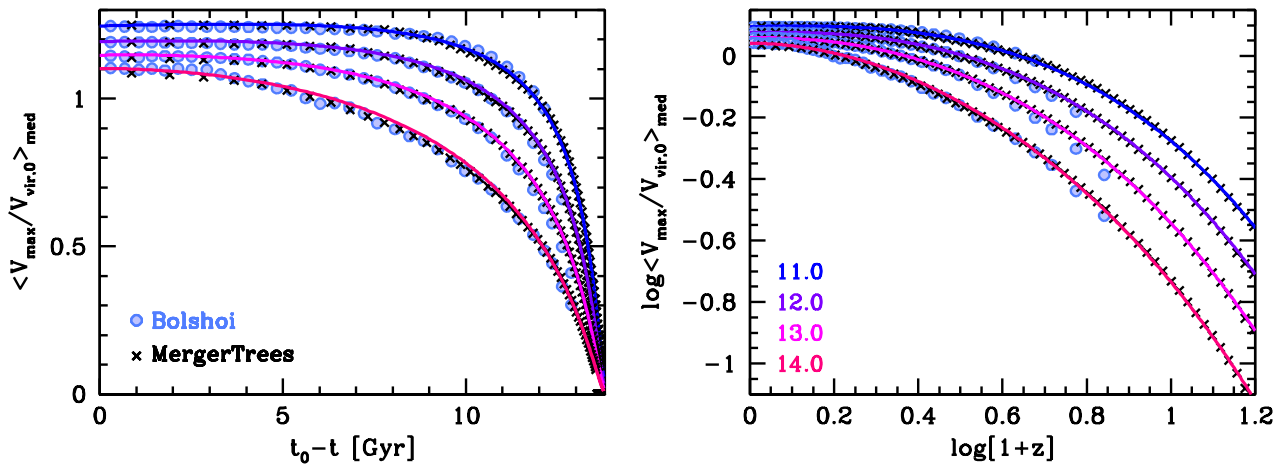,width=0.95\hdsize}}
\caption{Median PWGHs, $\langle V_{\rm max}/V_{{\rm vir},0}\rangle$ as
  function of lookback time (left-hand panel) and redshift (right-hand
  panel). Blue dots and black crosses are the results obtained from
  the Bolshoi simulation and MergerTrees, respectively, while the
  solid, colored lines are the predictions from our universal model
  described in \S\ref{sec:selfsim} and
  Appendix~\ref{App:model}. Results are shown for four different host
  halo masses, as indicated in the right-hand panel. Note the
  exquisite mutual agreement.}
\label{fig:fits}
\end{figure*}

Fig.~\ref{fig:fits} plots the median PWGHs for haloes of four different
host halo masses in the Bolshoi cosmology. Solid circles are the
results obtained from the Bolshoi simulation (using a halo mass bin
width of 0.2dex), while crosses are the results obtained using
MergerTrees. The solid, colored curves are the predictions for these
median PWGHs obtained using our universal model described above.  As is
evident, the model is in excellent agreement with the results from
MergerTrees, which in turn are in excellent agreement with the
simulation results. We have used MergerTrees to test the universal
model for halo masses and flat $\Lambda$CDM cosmologies spanning the
ranges $10^{9} \Msunh \lta M_0 \lta 10^{15} \Msunh$, $0.1 \leq
\Omega_{\rmm,0} \leq 0.5$, $0.6 \leq \sigma_8 \leq 1.0$, $0.9 \leq
n_\rms \leq 1.0$, and $0.6 \leq h \leq 0.8$. This amply covers the
range of cosmologies in agreement with current observations. For each
of those cases we find similar levels of agreement as shown in
Fig.~\ref{fig:fits}. In general, the model agrees with the predictions
from MergerTrees to better than a few percent.

Appendix~\ref{App:model} gives a detailed step-by-step description of
how to compute the average and/or median MAHs and PWGH using this
universal model. In addition, it also describes how this model can be
used to compute halo formation redshifts, $z_\rmf$, for any value of
$f$. 

\subsection{Mass Accretion Rates}
\label{sec:accrete}

The universal model for the MAHs presented above can also be used to
analytically compute the median or average mass accretion rates for
haloes of any mass, at any redshift, and for any $\Lambda$CDM
cosmology. Differentiating (\ref{unieq}) with respect to $\psi$ yields
that
\begin{equation}\label{dpsidt}
{\rmd \psi \over \rmd t} = {\partial \tomega \over \partial t} \,
\left[{\rmd \calF \over \rmd \psi} - 
{\partial \tomega \over \partial \psi} \right]^{-1}
\,,
\end{equation}
which, after some algebra, can be written in the form
\begin{equation}\label{dmdt}
\langle \dot{M} \rangle = {M \over t} {\calD \over \calH + \calS}\,,
\end{equation}
Here the angle brackets refer to either the median or the average,
depending on whether $\psi$ represents the median or the average, the
dot indicates the time derivative, and $\calD$, $\calH$, and $\calS$
are given by
\begin{equation}\label{calA}
\calD = {\partial {\rm ln} D \over \partial {\rm ln} t} \, \left[
{\delta_\rmc(t) \over \delta_\rmc(t) - \delta_\rmc(t_0)} - 0.004 \right]\,,
\end{equation}
\begin{equation}\label{calB}
\calS = \left\vert {\partial {\rm ln} \sigma \over \partial {\rm ln} M}\right\vert \, 
\left[
{\sigma^2(\psi M_0) \over \sigma^2(\psi M_0) - \sigma^2(M_0)} - 0.152 \right]\,,
\end{equation}
and
\begin{equation}\label{calD}
\calH = -{\rmd\ln\calF \over \rmd\ln\psi} = 0.4343 {a_2 a_3 \over 1 - a_2 \log\psi} +
{a_4 a_5 \psi^{a_4} \over 1 - \psi^{a_4}}
\end{equation}
Hence, $\calD$ describes the dependence on the linear growth rate,
$D(t)$, $\calS$ the dependence on the mass variance, $\sigma^2(M)$,
and $\calH$ is related to the functional form of the universal MAH.

\begin{figure*}
\centerline{\psfig{figure=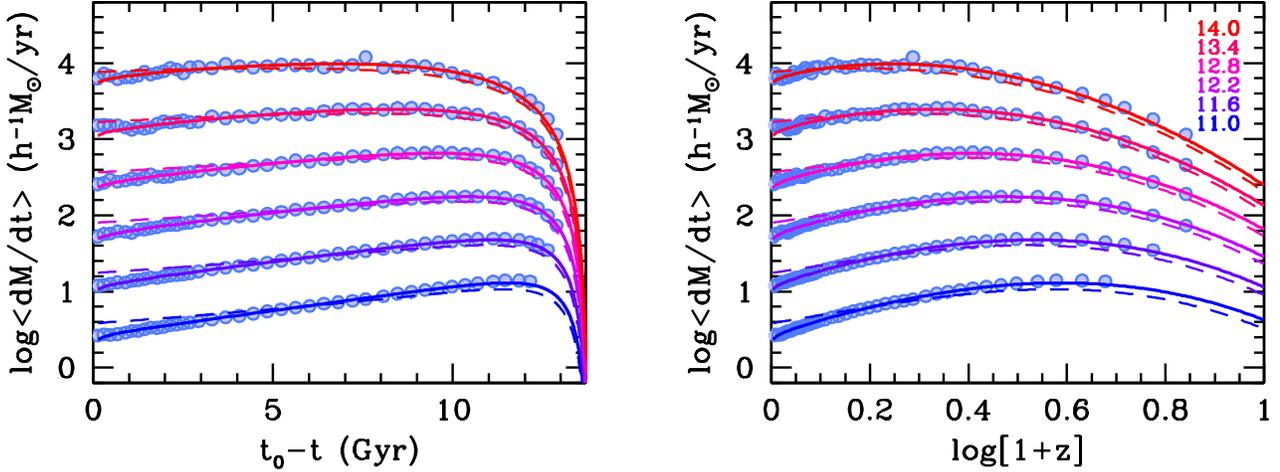,width=0.95\hdsize}}
\caption{Average halo accretion rates, $\langle \rmd M/\rmd t\rangle$,
  for the main progenitors of host haloes of different present day
  mass, $M_0$, as indicated.  Results are shown both as function of
  lookback time (left-hand panel) and as function of $\log[1+z]$
  (right-hand panel), to better accentuate the behavior at late and
  early times, respectively.  Solid circles are the results obtained
  from the Bolshoi simulation, by numerically differentiating the
  corresponding, average MAHs. The colored, solid curves are the
  predictions from our universal model (Eq.~[\ref{dmdt}]).  The
  colored, dashed curves are the corresponding model predictions of
  Fakhouri \etal (2010), given by Eq.~(\ref{Fak}), and are shown for
  comparison.}
\label{fig:accrate}
\end{figure*}

Fig.~\ref{fig:accrate} plots the average accretion rates for the main
progenitors of host haloes of different present day mass, $M_0$.
Solid circles are the results obtained from the Bolshoi simulation, by
numerically differentiating the corresponding, average MAHs. The
colored, solid curves are the model predictions computed using
Eq.~(\ref{dmdt}). Results are shown both as function of lookback time
(left-hand panel) and as function of $\log[1+z]$ (right-hand panel),
to better accentuate the behavior at late and early times,
respectively. As is evident, the model predictions are in excellent
agreement with the simulation results, providing further support for
our universal model.

The dashed curves are the model predictions of Fakhouri \etal (2010),
which are given by
\begin{eqnarray}\label{Fak}
\langle \dot{M} \rangle & = & 47.6\, h^{-1}\Msun {\rm yr}^{-1} 
\left({M \over 10^{12} h^{-1}\Msun}\right)^{1.1} (1 + 1.11z) \nonumber \\
& & \times \sqrt{\Omega_{{\rm m},0} (1+z)^3 + \Omega_{\Lambda,0}}\,.
\end{eqnarray}
These slightly underpredict the mass accretion rates at high $z$ and
overpredict $\langle \dot{M} \rangle$ at late times. We emphasize,
though, that although Fakhouri \etal have explicitly written how
their average mass accretion rate depends on $\Omega_{{\rm m},0}$ and
$\Omega_{\Lambda,0}$, their results are strictly {\it only} valid for
the cosmology adopted for the Millennium simulation, which differs
somewhat from the Bolshoi cosmology. Indeed, using our model to
compute $\langle \dot{M} \rangle$ for the Millennium cosmology, we
obtain results that are in much better agreement with Eq.~(\ref{Fak}),
but only at large redshifts; the discrepancy at late times ($t_0-t
\lta 3$Gyr) remains virtually unaltered.  Hence, we conclude that the
latter most likely reflects a discrepancy that arises from the use of
different halo finders and different algorithms to construct halo
merger trees (see discussion in \S\ref{sec:motivation}).


\section{Conclusions}
\label{sec:conclusion}

We have presented a detailed study of how dark matter haloes assemble
their mass and grow their (central) potential well. For an NFW
profile, the latter is directly proportional to the square of the
maximum circular velocity, which is why we characterize halo growth
via the mass accretion histories (MAHs), $\langle M/M_0\rangle$, and
the potential well growth histories (PWGHs), $\langle V_{\rm
  max}/V_{{\rm vir},0}\rangle$. Surprisingly the latter have received
little attention in the literature to this date, despite the fact that
$\Vmax$ has the advantange that it can be more reliably and robustly
measured than halo mass (both in simulations and in real data), and is
defined without ambiguity, freeing it from issues such as
`pseudo-evolution' that hamper interpretations of the MAHs. In
addition, $\Vmax$ is often used as the halo parameter of choice in
abundance matching, suggesting that it may be a better `regulator' of
galaxy formation than halo mass.

We have used results from both the large Bolshoi simulation, as well
as from merger trees constructed using the Parkinson \etal (2008)
algorithm. We have supplemented the latter with a method, developed by
Jiang \& van den Bosch (2014b), to compute the maximum circular
velocity, $\Vmax$, of the main progenitor as function of time. This
method uses the universal model between halo concentration and halo
formation time developed by Zhao \etal (2009).  We have demonstrated
that both methods yield results that are overall in excellent
agreement, both in terms of the average or median as well as in terms
of the scatter. However, we also identified a few small
inconsistencies. First of all, MergerTrees somewhat underpredict the
halo-to-halo variance at late times (Fig.~\ref{fig:compMAH}). Most
likely this is a manifestation of the fact that host haloes in
simulations can loose mass due to `ejected subhaloes', something that
is not accounted for in MergerTrees. Since most of this `lost mass'
remains bound to the host halo (and will re-accrete at some later
time), we do not consider this a particular failure of MergerTrees,
but rather a complication associated with how to assign mass to dark
matter haloes. Secondly,in order to match the slope of the
concentration-mass relation, we had to slightly modify the
concentration-formation time relation of Zhao \etal (2009). This is
most likely a consequence of the fact that Zhao \etal used a different
definition for main progenitor than adopted here.  Hence, our
concentration-formation time relation presented here
(Eq.~[\ref{concZhaomod}]) may be considered a recalibration of the
Zhao \etal model for cases in which the main progenitor is defined as
the most-massive, rather than the most-contributing, progenitor.
Finally, the Bolshoi simulation reveals distributions of halo
concentration at fixed halo mass that deviate from log-normal, in that
they have an extended tail towards low-concentration haloes. Such a
tail is absent in the distributions predicted using MergerTrees, and
is due to unrelaxed haloes. As a consequence, MergerTrees slightly
underpredicts the halo-to-halo variance in the PWGHs, but the effect
is small.

In agreement with numerous previous studies, we find that more massive
haloes assemble later. This not only holds for how they assemble their
mass, but also for how they build their central potential well. We
show that the latter precedes the former, illustrating the inside-out
build-up of dark matter haloes. We show that the haloes follow a tight
relation between $M(t)/M_0$ and $V_{\rm max}(t)/V_{{\rm max},0}$,
which is basically a manifestation of the fact that dark matter haloes
follow a universal density profile. As a rule of thumb, {\it the
  maximum circular velocity of a halo's main progenitor is already
  half the present day value by the time it has accreted only about 2
  percent of its final mass}. Consequently, dark matter haloes rapidly
grow their central potential, at early times, followed by an extensive
period in which the central potential deepends only very
slowly. During this `quiescent era' the halo mainly grows its
outskirts (see also Li \etal 2007).
   
In addition to a comparison with the results from the Bolshoi
simulation, we have also compared the predictions of MergerTrees with
the universal models for the median MAHs of Zhao \etal (2009) and
Giocoli \etal (2012).  We find that all models perfectly agree with
each other, and with the Bolshoi simulation results, at low redshift
($z \lta 1.5)$.  At higher $z$, the models of Zhao \etal (2009) and
Giocoli \etal (2012) slightly underpredict the median MAHs compared to
Bolshoi and to MergerTrees. Motivated by these findings we have
developed a new, universal model for both the average and median MAH,
which can also be used to predict the corresponding PWGH.  The model is
motivated by the fact that simulations and EPS-based merger trees have
shown that the unevolved subhalo mass function (i.e., the mass
function of haloes accreted directly by the main progenitor) is
universal. This means that, statistically speaking, all haloes grow in
mass by accreting the same haloes when normalized in mass by that of
the final host halo. Hence, the average (and median) MAHs for haloes
of different mass and/or different cosmology only differ from each
other because they accrete those progenitors at different times. This
suggests that MAHs should have a universal form when expressed in
terms of a `universal time'. We have found this universal time to be
given by
\begin{equation}\label{omegafmodfull}
\tomega(z|M,M_0,z_0) = {\delta_\rmc(z) - \delta_\rmc(z_0) \over
  \sqrt{\sigma^2(M) - \sigma^2(M_0)}} \, G^{0.4}(M,M_0,z_0)
\end{equation}
with $G(M,M_0,z_0)$ the perturbing function used in the Parkinson
\etal (2008) method to define the progenitor mass function.  We have
shown that, when plotted as function of $\tomega$, the average and
median MAHs have a universal (parametric) form given by $\calF(\psi) =
\tomega(\psi,z)$. Here $\psi$ is shorthand for either the average or
the median MAH, and $\calF(\psi)$ is a universal function that is
accurately fit by Eq.~(\ref{fitfunc}) with the best-fit parameters
listed in Table~1. As described in Appendix~\ref{App:model}, this
universal model for the MAH can also be used to compute the average or
median PWGH. We have tested this new, easy-to-use, universal model,
against the Bolshoi simulation results and against the predictions
based on MergerTrees, and found it to predict MAHs and PWGHs with
percent level accuracy over the entire range of halo masses and
$\Lambda$CDM cosmologies tested (see \S\ref{sec:selfsim} for details).

The fact that the unevolved subhalo mass function is universal not
only implies an universal MAH, it also implies that {\it the entire
  halo merger tree is invariant}; when expressing all progenitor
masses in units of the final host halo mass, and all times in term of
$\tomega$, one obtains a universal merger tree that can be used to
represent the merger tree for a halo of any mass in any ($\Lambda$CDM)
cosmology. This has important applications. For instance,
semi-analytical models for galaxy formation no longer have to use
large samples of merger trees to properly sample the entire population
of host haloes; rather, they can read in a sample of $\sim 100$ merger
trees for a halo of one particular mass (to sample the halo-to-halo
variance), and rescale these to represent the merger trees for any
other halo mass. In addition, changing cosmological parameters no
longer requires running a new set of merger trees, as the existing set
is trivially rescaled.

Finally, differentiating the universal MAH, we obtain a universal,
fully analytical model for the average (or median) mass accretion rate
of dark matter haloes, which is in excellent agreement with results
from the Bolshoi simulation. Unlike the fitting formula presented in
Neistein \& Dekel (2008a,b), Dekel \etal (2009), Genel \etal (2009),
McBride \etal (2009) and Fakhouri \etal (2010), all of which are only
valid for the particular cosmology adopted in the Millennium
simulation, this universal accretion rate is valid for any $\Lambda$CDM
cosmology.

We conclude by emphasizing that our results demonstrate the utility of
semi-analytical modeling in an era of research that increasingly
relies upon numerical results taken directly from simulations.
Semi-analytical models, such as that presented here, allow us to probe
a large dynamic range in mass and to explore cosmological dependencies
that would otherwise be excessively expensive in terms of CPU
requirements. In addition, semi-analytical models, by resorting to
simplified prescriptions of the underlying dynamics, are extremely
useful in gaining insight and understanding. The universal model
developed here is a clear example of the power of such an approach.

\section*{Acknowledgments}

We are grateful to all the people that have been involved with running
and analyzing the Bolshoi simulation for making their halo catalogs
and merger trees publicly available. Special thanks go out to Anatoly
Klypin and Peter Behroozi who have provided essential assistance with
handling and interpreting these magnificent data products.  APH is
supported by a fellowship provided by the Yale Center for Astronomy \&
Astrophysics. DFW is supported by the National Science Foundation
under Award No. AST-1202698.



\appendix


\section{Ejected Haloes}
\label{App:eject}

\begin{figure*}
\centerline{\psfig{figure=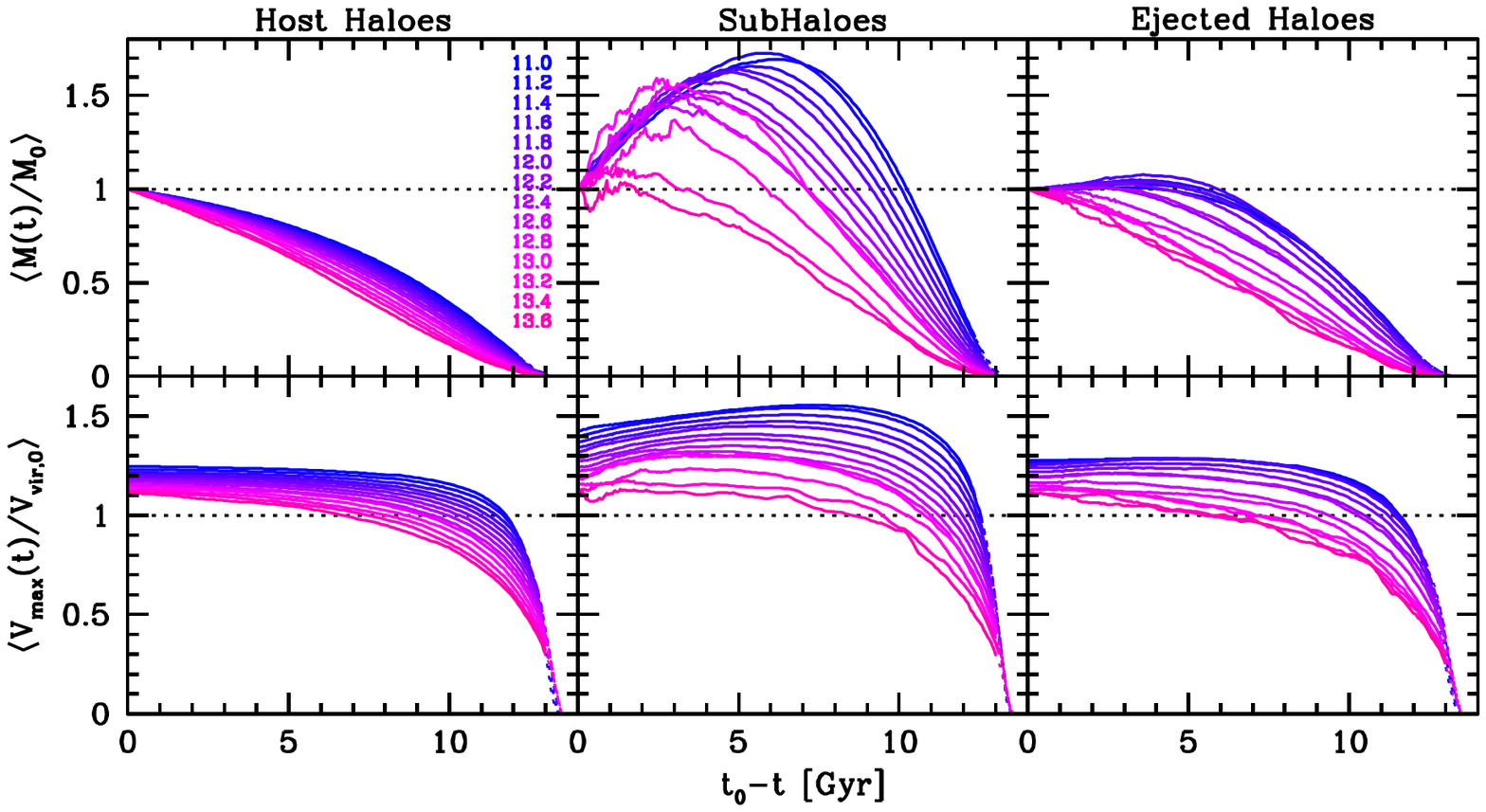,width=0.87\hdsize}}
\caption{Average MAHs (upper panels) and PWGHs (lower panels) as
  function of lookback time for host haloes (left), subhaloes (middle)
  and ejected host haloes (right). Different colors correspond to
  different present-day halo mass, as indicated in the upper left-hand
  panel. Note that these different classes of haloes have clearly
  distinct MAHs and PWGHs. Hence, when studying the behavior of host
  haloes in numerical simulations, it is prudent to remove subhaloes
  and ejected host haloes from the sample.}
\label{fig:MAHeject}
\end{figure*}

Fig.~\ref{fig:MAHeject} compares the average MAHs obtained from the
Bolshoi simulation for host haloes (left-hand panels), subhaloes
(middle panels), and ejected haloes (right-hand panels). Once again
different colors correspond to different bins of $z=0$ halo mass (each
bin is 0.2dex wide), as indicated. We only plot results for mass bins
for which we have at least 20 haloes per category, which restricts
the comparison to haloes with $M_0 \leq 10^{13.7} \Msunh$.

We emphasize that it is not necessarily particularly meaningful to
average the MAH for subhaloes or ejected haloes in bins of present-day
mass; however, the main point of Fig.~\ref{fig:MAHeject} is to
demonstrate that subhaloes and ejected haloes have MAHs that differ
substantially from those of host haloes of the same present-day mass
and therefore have to be excluded from the samples of `host
haloes'. Although this is pretty obvious for subhaloes, it is less
clear for ejected host haloes, and virtual all studies to date of the
MAHs of dark matter haloes based on numerical simulations have
included ejected haloes in their samples. Although their fractional
contribution is small (see Fig.~\ref{fig:frac}), their average MAHs
are sufficiently different that they can still cause a mild (but
significant) distortion of the averages, especially for low mass
haloes. Motivated largely by the work of Wang \etal (2009b), Geha
\etal (2012), Wetzel \etal (2014), and others, we believe that ejected
haloes are more akin to subhaloes than to host haloes, and we
therefore remove them from our sample of host haloes.


\section{Cosmology Dependence}
\label{App:cosmo}

As discussed in \S\ref{sec:comparison}, various studies have presented
fitting functions for the average and/or median MAHs (or their
time-derivatives) based on one particular numerical simulation (e.g.,
Neistein \& Dekel 2008a; McBride, Fakhouri \& Ma 2009; Genel \etal
2009; Fakhouri \etal 2010; Wu \etal 2013). In this Appendix, we
demonstrate that the average MAHs depend appreciably on cosmology, and
that these fitting functions are therefore only applicable for the
cosmology used to run the numerical simulation.

\begin{figure*}
\centerline{\psfig{figure=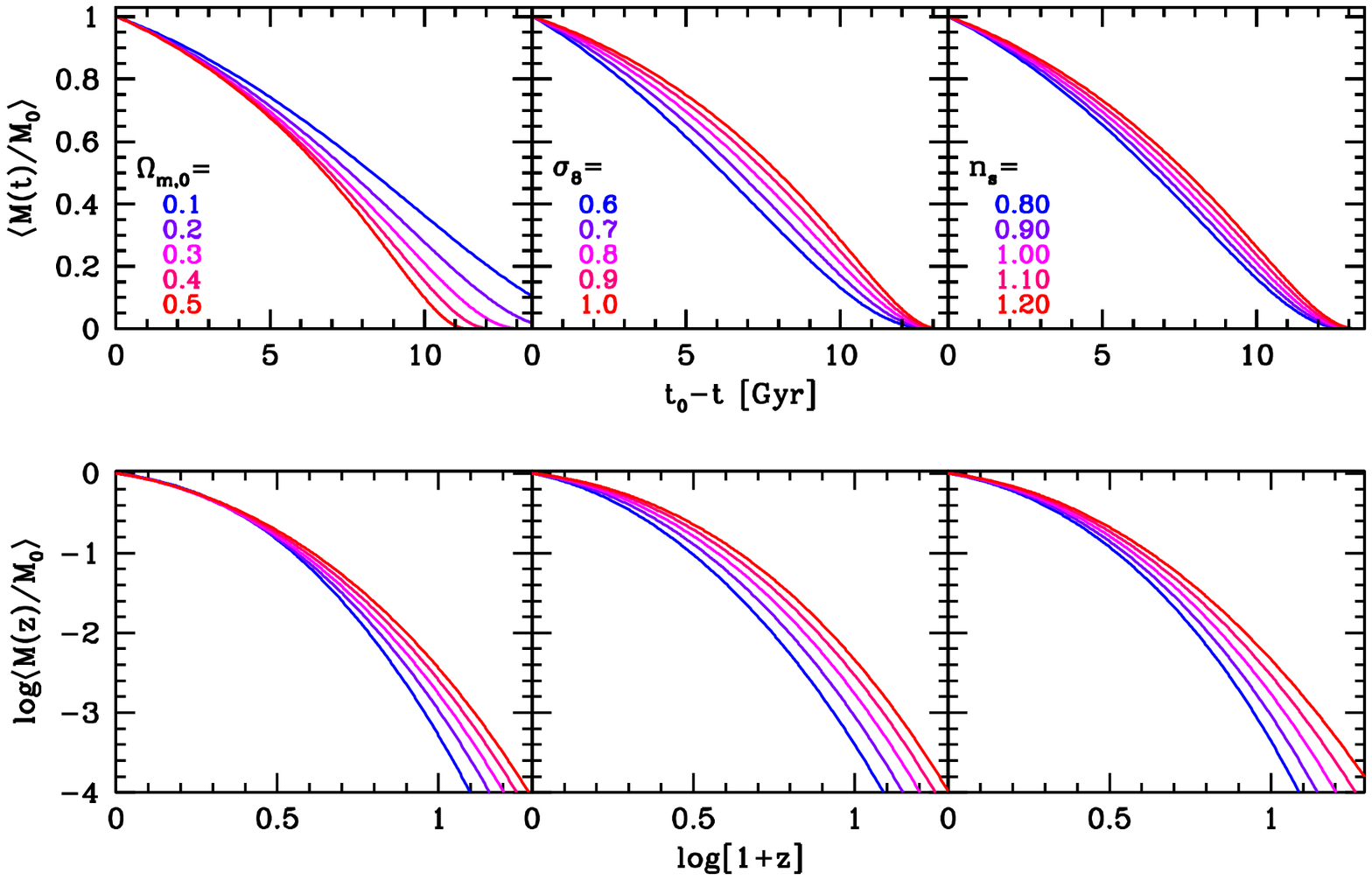,width=0.87\hdsize}}
\caption{Average MAHs for haloes of mass $M_0 = 10^{13}\Msunh$ in
  different, flat $\Lambda$CDM cosmologies.  Upper and lower panels
  plot $\langle M(t)/M_0 \rangle$ as function of lookback time and
  $\log\langle M(z)/M_0 \rangle$ as function of $\log[1+z]$.
  Different colors correspond to different cosmologies, where only one
  cosmological parameter is varied at a time. In the left-hand panel
  this is $\Omega_{\rm m,0} = 1-\Omega_{\Lambda,0}$, which is varied
  from 0.1 to 0.5, in the middle panel $\sigma_8$ is varied from 0.6
  to 1.0, and in the right-hand panel the spectral index, $n_\rms$, is
  varied from $0.8$ to $1.2$. See text for a detailed discussion.}
\label{fig:cosmo}
\end{figure*}

Fig.~\ref{fig:cosmo} plots the average MAHs for a halo of mass $M_0 =
10^{13}\Msunh$ in different, flat $\Lambda$CDM cosmologies. Each
average MAH is obtained averaging over 2000 realizations of
MergerTrees.  Upper and lower panels plot $\langle M(t)/M_0 \rangle$
as function of lookback time and $\log\langle M(z)/M_0 \rangle$ as
function of $\log[1+z]$, accentuating the behaviors at late and early
times, respectively.  The base-cosmology has $\Omega_{\rm m,0} = 1 -
\Omega_{\Lambda,0} = 0.3$, $\Omega_{\rm b,0} = 0.045$, $h = 0.7$,
$\sigma_8 = 0.8$ and $n_\rms = 1.0$, and is represented by the magenta
curves.  The other curves are the average MAHs, for haloes of the same
present-day mass, in cosmologies in which only one parameter is
changed with respet to this base cosmology: in the left-hand panel we
change $\Omega_{\rm m,0}$ from $0.1$ to $0.6$ (note that we change
$\Omega_{\Lambda,0}$ as well, to assure a flat geometry), in the
middle panel we change $\sigma_8$ from 0.6 to 1.0, and in the
right-hand panel we change the spectral index, $n_\rms$, from 0.8 to
1.2.

Changing the cosmological matter density, $\Omega_{\rm m,0}$, changes
the linear growth rate.  As a result, haloes in low-$\Omega_{\rm m,0}$
cosmologies assemble earlier than in a high-$\Omega_{\rm m,0}$
cosmology, when expressed in terms of lookback time. Note, though,
that this trend is reversed when plotted as function of redshift, due
to the fact that changes in $\Omega_{\rm m,0}$ also affect the
expansion history, and thus the time-redshift relation. Increasing the
normalization of the power spectrum, $\sigma_8$, boosts the amplitudes
of the density perturbations, causing structure to grow earlier. And
finally, increasing the spectral index, $n_{\rm s}$, boosts the power
on small scales relative to large scales, causing haloes to assemble
earlier. Without showing the results, we emphasize that the magnitude
of this $n_\rms$-dependency depends quite strongly on halo mass; it is
significantly stronger for low mass haloes with $M_0 = 10^{11} \Msunh$
and becomes negligible for massive clusters with $M_0 = 10^{15}
\Msunh$.

All in all, it is clear from Fig.~\ref{fig:cosmo} that changes in the
cosmological parameters of order 10 percent have a non-neglible impact
on the MAHs of dark matter haloes. This motivates the development of
universal models, such as that presented in this paper, rather than
the use of fitting functions that are only valid for a single
cosmology. For the sake of brevity, we do not show how changes in
cosmological parameters impact the PWGHs, but this is easily assessed
using the Universal model described in \S\ref{sec:uni} and
Appendix~\ref{App:model}.
 

\section{Recipe for computing MAH and PWGH}
\label{App:model}

This Appendix details how to compute the average and/or median MAH and
PWGH for a host halo of mass $M_0$ at redshift $z_0$ in a $\Lambda$CDM
cosmology.

As described in \S\ref{sec:uni}, computing the average or median
MAH reduces to numerically solving 
\begin{equation}\label{root}
\calF(\psi) = \tomega(\psi,z)
\end{equation}
for $\psi$ at given $z$, or $z$ at given $\psi$. Here
$\tomega(\psi,z)$ is the universal `time-coordinate' given by
Eq.~(\ref{omegafmod}) with $\gamma =0.4$, and $\calF(\psi)$ is a
fitting function (Eq.~\ref{fitfunc}) that describes the universal,
parametric relation between $\psi$ and $\tomega$. The values for the
corresponding parameters are listed in Table~1, and depend on whether
$\psi$ represents the average or the median. Note that solving
Eq.~(\ref{root}) for $z$ for a given value of $\psi$ is equivalent to
computing the formation redshift $z_\rmf$ for $f = \psi$. Hence,
Eq.~(\ref{root}) can be used to compute the mean or median halo
formation redshifts, $z_\rmf$, for any value of $f$.  This is similar
to the model developed by G12, except that our model is valid for main
progenitors being defined as the most-massive progenitors, whereas the
G12 model corresponds to main progenitors being defined as the
most-contributing progenitors (see \S\ref{sec:SIM}).

The following step-by-step procedure outlines how to compute
the average (or median) MAH and PWGH:
\begin{enumerate}
\item Define a vector $z_i$ ($i=1,2,...,N$) with the redshifts at
  which to compute the average or median MAH.

\item For each $z_i$, use a root-finder to solve Eq.~(\ref{root}) for
  $\psi$, and store the resulting values in a vector $\psi_i$

\item For each $z_i$, use interpolation of $\psi_i$ to find the
  corresponding $z_{0.04,i}$, defined by $\psi(z_{0.04,i}) =
  0.04\,\psi(z_i) = 0.04 \psi_i$.

\item For each $z_i$ and $z_{0.04,i}$ compute the corresponding proper
  times, $t_i$ and $t_{0.04,i}$, and use Eq.~(\ref{concZhaomod}) to
  compute the halo concentration, $c_i$, of the main progenitor at
  $z_i$.

\item Use Eqs.~(\ref{VmaxHost}) and~(\ref{Vvir}) to compute
  the corresponding PWGH, $V_{{\rm max},i}/V_{{\rm vir},0}$.

\end{enumerate}

A simple Fortran code that computes the average and median MAHs, PWGHs,
mass accretion rates, and main progenitor concentrations as function
of redshift is available for
download\footnote{http://http://www.astro.yale.edu/vdbosch/PWGH.html}.
For a given cosmology, the program takes only a few seconds, on a
regular desktop computer, to compute these quantities for tens of halo
masses using $N = 400$.

\label{lastpage}

\end{document}